\documentclass[acmtog, authorversion]{acmart}
\AtBeginDocument{%
  \providecommand\BibTeX{{%
    \normalfont B\kern-0.5em{\scshape i\kern-0.25em b}\kern-0.8em\TeX}}}

\copyrightyear{2024}
\acmYear{2024}
\setcopyright{rightsretained}
\acmConference[Authorversion]{}{2024}{}
\acmISBN{978-1-4503-XXXX-X/X/XX}

\usepackage{color, colortbl}
\usepackage{multirow}
\usepackage{subfig}
\usepackage{longtable}
\usepackage{hyperref}
\usepackage[font=small,labelfont=bf]{caption}
\usepackage{wasysym}
\usepackage[nolist,nohyperlinks, printonlyused, withpage]{acronym}
\usepackage{enumitem}

\definecolor{cred}{RGB}{198, 39, 55}
\definecolor{cyellow}{RGB}{218, 163, 0}
\definecolor{cgreen}{RGB}{0, 150, 122}

\begin{document}

\title[Finding a Way Through the Social Media Labyrinth]{Finding a Way Through the Social Media Labyrinth: Guiding Design Through User Expectations}

\author{Thomas Mildner}
\affiliation{%
  \institution{University of Bremen}
  \city{Bremen}
  \country{Germany}
}
\email{mildner@uni-bremen.de}
\orcid{0000-0002-1712-0741}

\author{Gian-Luca Savino} {
\affiliation{%
  \institution{University of St.Gallen}
  \city{St.Gallen}
  \country{Swiss}
}
\email{gian-luca.savino@unisg.ch}
\orcid{0000-0002-1233-234X}
}

\author{Susanne Putze}
\affiliation{%
  \institution{University of Bremen}
  \city{Bremen}
  \country{Germany}
}
\email{sputze@uni-bremen.de}
\orcid{0000-0002-3072-235X}

\author{Rainer Malaka}
\affiliation{%
  \institution{University of Bremen}
  \city{Bremen}
  \country{Germany}
}
\email{malaka@tzi.de}
\orcid{0000-0001-6463-4828}

\renewcommand{\shortauthors}{Mildner et al.}

\begin{abstract}
Social networking services (SNS) have become integral to modern life to create and maintain meaningful relationships. Nevertheless, their historic growth of features has led to labyrinthine user interfaces (UIs) that often result in frustration among users -- for instance, when trying to control privacy-related settings. This paper aims to mitigate labyrinthine UIs by studying users' expectations ($N=21$) through an online card sorting exercise based on 58 common SNS UI features, teaching us about their expectations regarding the importance of specific UI features and the frequency with which they use them. Our findings offer a valuable understanding of the relationship between the importance and frequency of UI features and provide design considerations for six identified UI feature groups. Through these findings, we inform the design and development of user-centred alternatives to current SNS interfaces that enable users to successfully navigate SNS and feel in control over their data by meeting their expectations.
\end{abstract}

\ccsdesc[500]{Human-centered computing~Empirical studies in HCI}
\ccsdesc[300]{Human-centered computing~Empirical studies in interaction design}
\ccsdesc[100]{Human-centered computing~Interaction design theory, concepts and paradigms}
\ccsdesc[300]{Security and privacy~Usability in security and privacy}

\keywords{SNS, social media, deceptive design, dark patterns, ethical user interfaces, ethical design, user experience, user expectation, card sorting, user-centered design}
\maketitle

\textcolor{red}{\textbf{Draft: May 12, 2024}}

\section{Introduction}
Social networking services (SNS) are ubiquitous in many people's everyday lives as both personal and professional drivers for maintaining relationships, retrieving information, and engaging in communities~\cite{stone_why_2022}. Yet, in spite of their success, the experience of SNS users is not entirely positive. Misaligned expectations, unfulfilled satisfactions~\cite{mildner_ethical_2021, schaffner_understanding_2022}, and the feeling of losing control over one's personal data~\cite{schoenebeck_giving_2014,ernala_how_2020} decrease users' satisfaction when using related platforms. To a certain degree, this is the result of difficult-to-navigate user interfaces (UIs), particularly settings menus~\cite{mildner_about_2023,gunawan_comparative_2021}, leading to increasing demands for better control over personal data~\cite{wang_i_2011,lyngs_2020,alemany_review_2023}.

Noticeably, two main reasons factor into the difficulty of SNS navigation and, thus, a loss-of-control feeling among their users.
Firstly, SNS have matured into complex applications with a wide range of features to engage with others and settings users have to maintain~\cite{mildner_ethical_2021,dhingra_historical_2019,mcintyre_evolution_2014}. Many SNS feature multiple feeds or timelines, options for public or personal discourse, and a wide range of controls to customise user experience, such as, but not limited to, control of personal data, advertisement-related data, and notifications from SNS applications. 
Secondly, research has identified a host of dark patterns in SNS~\cite{mildner_about_2023,mildner_defending_2023,schaffner_understanding_2022,habib_identifying_2022} as well as in the design of many of these individual features.
In the context of SNS, dark patterns, also referred to as deceptive design patterns~\footnote{We chose to use the term ``dark pattern'' consistent with prior research efforts to describe design strategies that obfuscate or hide consequences of (online) interfaces. Nonetheless, we recognise that the ACM Diversity, Equity, and Inclusion Council has recently~\cite{acm_words_2023} identified the term as contentious due to possible negative implications when implying evilness or malintent. Our usage of the term adheres to its initial definition, highlighting the hidden nature of these design strategies with concealed consequences on users~\cite{brignull_deceptive_2023}. We further acknowledge that this discourse is ongoing and that there is currently no alternative term that fully encapsulates the deceptive, manipulative, obstructive, or coercive nature inherent to the original term.}, are used to steer users' attention~\cite{monge_roffarello_defining_2023} or unwillingly increase their engagement while governing their decisions~\cite{mildner_about_2023}. Building on the existing taxonomy of dark patterns captured and described in related work~\cite{gray_ontology_2023,gray_ontology_2024}, a recent effort by Mildner et al.~\cite{mildner_about_2023} identified more than 40 different dark patterns across popular SNS platforms, while Monge Rofarello et al.~\cite{monge_roffarello_defining_2023} described design strategies on SNS aimed to capture their users' attention. Furthermore, in a different work, Mildner et al.~\cite{mildner_defending_2023} illustrated SNS users' difficulty in effectively protecting themselves from these strategies, which is in line with similar prior studies~\cite{digeronimo2020,bongard-blanchy_i_2021}. At least partially responsible for users' struggles in this regard, the concept of \textit{Labyrinthine Navigation}~\cite{mildner_about_2023} describes tangled UI structures that hinder users from successfully and effectively navigating SNS interfaces when controlling personal data within and outside of particular SNS platforms.
This combination of increasingly complex interfaces and the prevalence of dark patterns have led to SNS UIs where settings --- containing crucial elements to control personal data --- are hidden deep within complex and nested interface structures. Although we cannot know whether such design choices fall under malintent, their users experience the consequences through bad usability and a lack of control.

In this research, we take a first step towards improved usability and control for SNS users by understanding their expectations regarding SNS interface design and complexity. To this end, we consider a user-centred design (UCD) approach for investigating the importance of relevant SNS features and the frequency with which they are used. We analysed Facebook's interface as a prototypical example for SNS UIs, as it remains highly relevant after almost two decades in service. Facebook has long been the focus of privacy-concerned research~\cite{wang_privacy_2013,habib_identifying_2022,prajwal_examining_2022} and has been critically reviewed repeatedly from the eyes of public media outlets~\cite{singer_why_2018,social_2020}. Through constantly changing and extending its features, Facebook presents particular challenges for its users to keep personal settings in their desired states and navigate the interface successfully. 

In our study, we conducted a card sorting experiment based on Facebook's interface with 21 participants identifying both important and frequently used interface elements as well as seemingly unimportant and less used ones to gain insights about users' expectations about individual and groups of SNS features. Through this study, we aim to answer the following research question:

\begin{itemize}
    \item [\textbf{RQ:}] 
    What specific design considerations should be taken into account to align SNS UIs with their users' expectations?
\end{itemize}

Based on our results, we discuss design considerations to structure SNS UI features according to our participants' ratings in terms of the importance and frequency in which a feature is used. 
Moreover, we identified six groups of SNS UI elements that further capture users' expectations and offer an initial hierarchy within SNS UI features: (1) ``User Support'', (2) ``Legal \& Policy Compliance'', (3) ``Data Security \& Privacy'', (4) ``Profile \& Account Management'', (5) ``Visibility Control'', and (6) ``User Experience Customization''. These groups include common functionalities implicit in SNS but also less used features for privacy, security, and control over users' data. In tandem, these insights offer opportunities to rethink and restructure current SNS to avoid labyrinthine UI structures and instead aid users in navigating them successfully to maintain features and settings according to their preferences.
In contrast to current design efforts of commercial SNS applications, which deal with a large number of features and are affected by dark patterns, this paper argues solely from the users' perspective and their expectations. We acknowledge that designing good SNS applications in terms of UI and settings menus is a challenging task. With our design considerations, we contribute a first step towards improving the status quo. 


\section{Related Work}
The main focus of this research lies within traditional UCD concepts in the context of SNS, intending to understand users' perceptions to design optimal UIs that respond to their expectations. However, our study draws from recent efforts in HCI spotlighting unethical, exploitative design strategies that decrease users' ability to make informed decisions through deceptive and manipulative dark patterns. The related work begins by highlighting traditional user-centred and ethical design. Afterwards, we continue with SNS-related studies identifying problematic design and dark patterns that limit user agency.

\subsection{From User-Centred Design to Deception}
Traditionally, HCI provides designers with the means to develop user-centred interfaces that should be intuitive and easy to navigate. 
Extending core principles of UCD, the ethically driven school of Value Sensitive Design (VSD)~\cite{doorn_value_2013} promotes the necessity to uphold users' autonomy and make consequences of interactions transparent to the user. While VSD promotes user autonomy, other interfaces are designed to guide users through complex interactions. In this regard, nudges~\cite{thaler_nudge_2008} and persuasive design~\cite{fogg_persuasive_2009} can be deployed to increase usability in terms of efficiency and engagement, but not necessarily transparency. Although these concepts find many useful applications, especially in health-related contexts~\cite{alexandrovsky_serious_2021}, they can be exploited to undermine users' ability to make informed decisions~\cite{hansen_nudge_2013,chen_practitioners_2022} leading to deceptive or manipulative interfaces~\cite{brignull_deceptive_2023,gray_ontology_2023}. Offering some insight into how design can accommodate user autonomy to avoid deceptions, Leimstädter ~\cite{leimstadtner_investigating_2023} build on Hansen and Jespersen's framework ~\cite{hansen_nudge_2013} to promote reflection through design friction, however, at the cost of user experience and restricted usability. In the scope of persuasive technologies, recent work by Bennett et al.~\cite{bennett_how_2023} has underlined the general relevance of user agency and autonomy. However, the authors notice certain ambiguities in the terms' usage in related work. In this paper, we follow traditional UCD concepts and Bennett et al.'s terminology suggestion to study users' expectations when interacting with SNS features in terms of importance and frequency~\cite{bennett_how_2023}. 

\subsection{User Agency in SNS}
Similar to autonomy-related work, within the HCI peripheral, research has investigated the effects of SNS on user agency for some time now~\cite{jeong_what_2016,schaffner_understanding_2022} -- particularly regarding users' privacy behaviour~\cite{kokolakis_privacy_2017,andalibi_human_2020,alemany_review_2023}. The lack of agency to use SNS as desired may place users in a vulnerable spot. In this regard, an array of studies illustrate the contrast between the positive effects of SNS increasing social connectedness~\cite{ahn2013social,sinclair2017facebook} and misuse of SNS, leading to negative consequences on users' well-being~\cite{christakis2010internet,twenge2018,coyne2020,beyens2020effect}. These tensions highlight a continuous need to study how SNS affect their users and what interface features are responsible for potentially problematic outcomes. 

This need is further amplified by the constant change and increase of SNS features~\cite{dhingra_historical_2019,mcintyre_evolution_2014}. Since their advent, SNS have grown into sophisticated platforms that extend their original features when offering users a wide variety of options to engage with content and other users~\cite{mildner_ethical_2021}. This upscale of features has led to complex UIs that users may find difficult to navigate~\cite{digeronimo2020,mildner_about_2023}. Based on a user study on YouTube's mobile interface, Lukoff et al.~\cite{lukoff_how_2021} noticed design strategies deployed by the platform limiting a sense of agency among its users and found opportune interface mechanisms that could increase their ability to use the platform as preferred. A common approach to enable users in this regard is the implementation of design interventions. Concerned with Facebook's interface, Lyngs et al.~\cite{lyngs_2020} developed two interventions that would either remind users about their usage goals or remove Facebook's newsfeed to help users not get distracted. Although their results contain limitations, they demonstrate the benefits of increased control over one's usage behaviour. In a similar vein, Masaki et al. \cite{masaki_exploring_2020} demonstrate how nudges used as design interventions can protect users from exposing personal information unwillingly. Their results are in line with prior results by Wang et al.~\cite{wang_effects_2014}, who confirm that positive impact interface nudges and friction design can have to help users reflect on their decisions before engagement. 
While related work presents important design interventions as countermeasures to problematic interface design that limit user agency, in this work, we aim to understand SNS users' expectations to inform UIs that avoid the implementation of problematic designs. 
To this end, we discuss design considerations for structuring UIs based on users' expectations regarding individual and groupings of SNS UI features. 

\subsection{SNS Breaking Users' Expectations}
Work focusing on design interventions suggests a misalignment of interests between providers of SNS and their users. This discrepancy could stem from commercial incentives~\cite{zuboff_surveillance_2023}, leading to unethical design, such as dark patterns, in SNS' UIs~\cite{chivukula_wrangling_2023,gray_ethical_2019}. Dark patterns are design strategies that prohibit users from making informed decisions by obfuscating or obstructing informed decision-making~\cite{brignull_deceptive_2023,gray_dark_2018}. As per their nature, dark patterns are difficult to avoid~\cite{digeronimo2020}, even when participants were made aware of their existence~\cite{bongard-blanchy_i_2021}, and SNS are not exempt from this~\cite{mildner_defending_2023}. 
Consequently, users may be unable to maintain account settings aligned with their preferences depending on the screen modality used to access a service~\cite{gunawan_comparative_2021,mildner_defending_2023} or feel restricted from deleting their accounts alltogether~\cite{schaffner_understanding_2022}. 

Arguably, dark patterns restrict users' agency and autonomy to use systems in terms of their beliefs or values and break their expectations. Recent work done by Mildner et al.~\cite{mildner_about_2023} dismantled popular SNS interfaces identifying a range of dark patterns based on a corpus of 80 types. Moreover, the work described five SNS-specific dark patterns that subscribe to engaging and governing strategies. The engaging strategies fall in line with designs that draw users' attention towards themselves, as described as attention-capturing by Monge Roffarello~\cite{monge_roffarello_defining_2023}. The governing strategies, on the other hand, convey interfaces that steer users' interactions while disregarding their goals. Motivated by the \textit{Labyrinthine Navigation} dark pattern falling under this strategy, describing complex and nested interfaces users easily get lost in, we recorded Facebook's interface to study users' perception of its features. Here, our aim was to learn about users' expectations in terms of the importance of features as well as the frequency with which users use them. Based on these criteria, we gained an in-depth understanding of how SNS UI features can be structured to accommodate user preferences.







\section{Method}
To understand SNS users' expectations of SNS UI features, we conducted a card sorting study with 21 participants, including 58 cards representing typical SNS features collected from Facebook's interface. The study was designed to be completed unsupervised and online through the web application Miro~\cite{realtimeboard_inc_miro_2023}. The online study setup was self-contained, meaning the instructions and the task were embedded on the Miro board. Thus, participants did not have to leave the platform throughout the exercise and could concentrate on completing the study.

\subsection{Selection of Cards}
As a basis for the cards, we analysed the interface of Facebook's mobile application. The decision fell on Facebook as it remains a popular SNS that, in the course of almost two decades, changed its UI and adopted different strategies, growing into the platform it is today. We screen-recorded a walkthrough of the complete Facebook application\footnote{We recorded usage based on Facebook version 397.0 on iOS.} and identified a total of 102 UI features, including the feed, profile page, and settings menu. In an attempt to limit the scope of cards to suit the purpose of this study (i.e. understanding expectations toward general SNS features), we excluded certain UI features that were specific to Facebook (e.g. marketplace or dating features) or exceeded the scope of this research (e.g. payment methods, management tools for professionals). Finally, this reduction resulted in 58 cards of relevant SNS UI features that participants were asked to sort.

\subsection{Card sorting Procedure}
\begin{figure*}[t!]
    \centering
    \includegraphics[width=\textwidth]{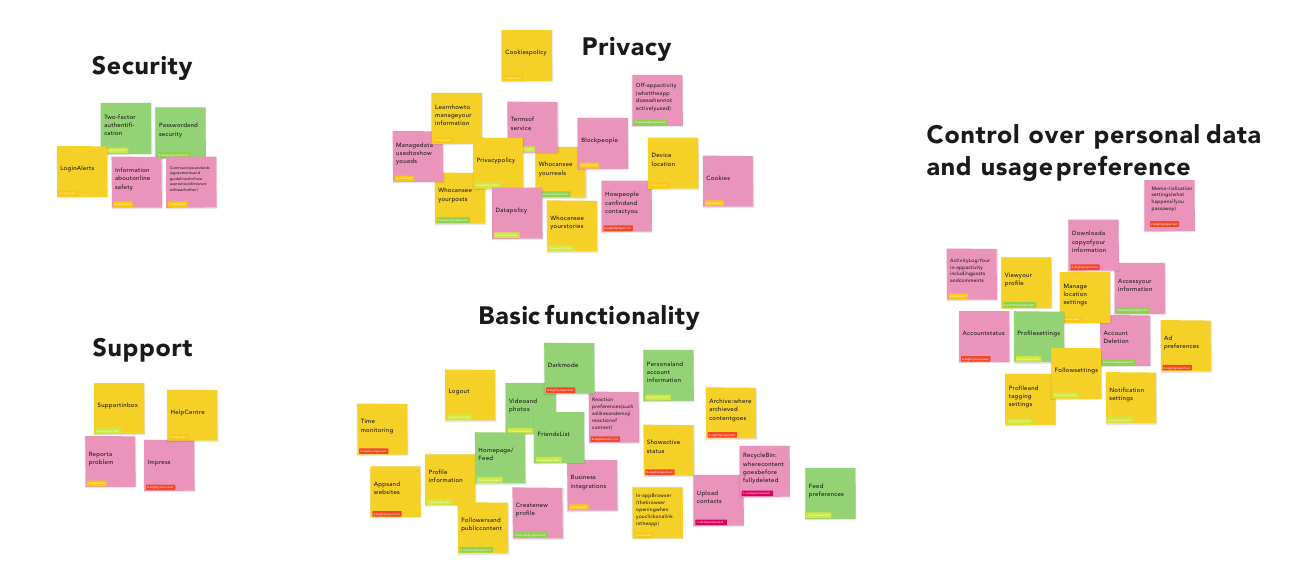}
    \caption{This figure displays one participant's card sorting results featuring 5 groups of cards with the labels security, privacy, support, basic functionality and control over personal data and usage preference. The cards are both colour-coded and include importance ratings according to the study design.}
    \Description{
    This figure displays one participant's card sorting results featuring 5 groups of cards with the labels security, privacy, support, basic functionality and control over personal data and usage preference. The cards are both colour-coded and include importance ratings according to the study design.
    }
    \label{fig:cardsorting}
\end{figure*}

Each participant was provided with the necessary information to participate in the study via email, together with a consent form and demographic questionnaire to fill out. After we received their consent and demographic data, participants were given a link to their individual Miro boards where the online card sorting task took place (see Figure~\ref{fig:cardsorting} for one participant's card sorting results). The study was designed to last about 40 minutes and involved three parts: First, participants were asked to follow a traditional card sorting approach by grouping cards based on similar traits to learn about the relatedness of UI features. Second, we asked them to assign each card an importance score from 1-5 (not important at all - extremely important), following a Likert scale. Lastly, participants colour-coded the cards depending on the frequency with which they would use a particular feature (``frequent'' usage in \textbf{\textcolor{cgreen}{green}}, ``moderately'' in \textbf{\textcolor{cyellow}{yellow}}, and ``rarely to never'' in \textbf{\textcolor{cred}{red}}). We acknowledge that such a 3-point Likert scale is restricting the analysis but follow advice~\cite{jacoby_1971_three} that it can produce interesting insights, especially when a study design demands participants' focus over longer periods of time. The data gained through these additional tasks informed us about the relevance of each feature based on two criteria --- importance and frequency --- which offered further insights into our participants' expectations for the UI features.

\subsection{Participant Demographics}
Participants were recruited through various university computer science and HCI programs. Participation was entirely voluntary, and participants were rewarded 10€ for taking part in the study. To qualify for this study, participants had to engage with social media on a weekly basis and needed to be enrolled in HCI-related programs. The latter requirement was chosen as we aimed for some sensibility towards HCI research and technological literacy. In total, 23 participants participated in this card sorting study. However, two participants had to be excluded for incomplete participation. The remaining data from 21 participants was therefore included in the further analysis. Of these 21 participants, nine self-identified as female and twelve as male. At the time of conducting the study, the participants' mean age was 26.52 years (sd=3.56). They were recruited from Germany ($n=10$), Netherlands ($n=1$), Switzerland ($n=8$), and the USA ($n=2$). Between participants, their highest education included a high-school diploma ($n=1$), bachelor's degree ($n=9$), and master's degree ($n=11$). They used SNS an average of 6.86 days per week ($sd=0.48$).

\section{Findings}
In this section, we first present the results of the grouping aspect of the card sorting task. We used hierarchical clustering based on a similarity matrix generated from each participant's card sorting results. This allows us to assess the relationship between individual cards and to create average groups based on the participants' individual decisions~\cite{capra_factor_2005}. We then turn to the individual features and report each UI feature's importance and frequency ratings. As the combined data from the 58 cards is too large to be presented in this paper, we focus on the most relevant findings with the complete data included in this paper's supplementary material.

\begin{figure*}[t!]
    \centering
    \includegraphics[width=\textwidth]{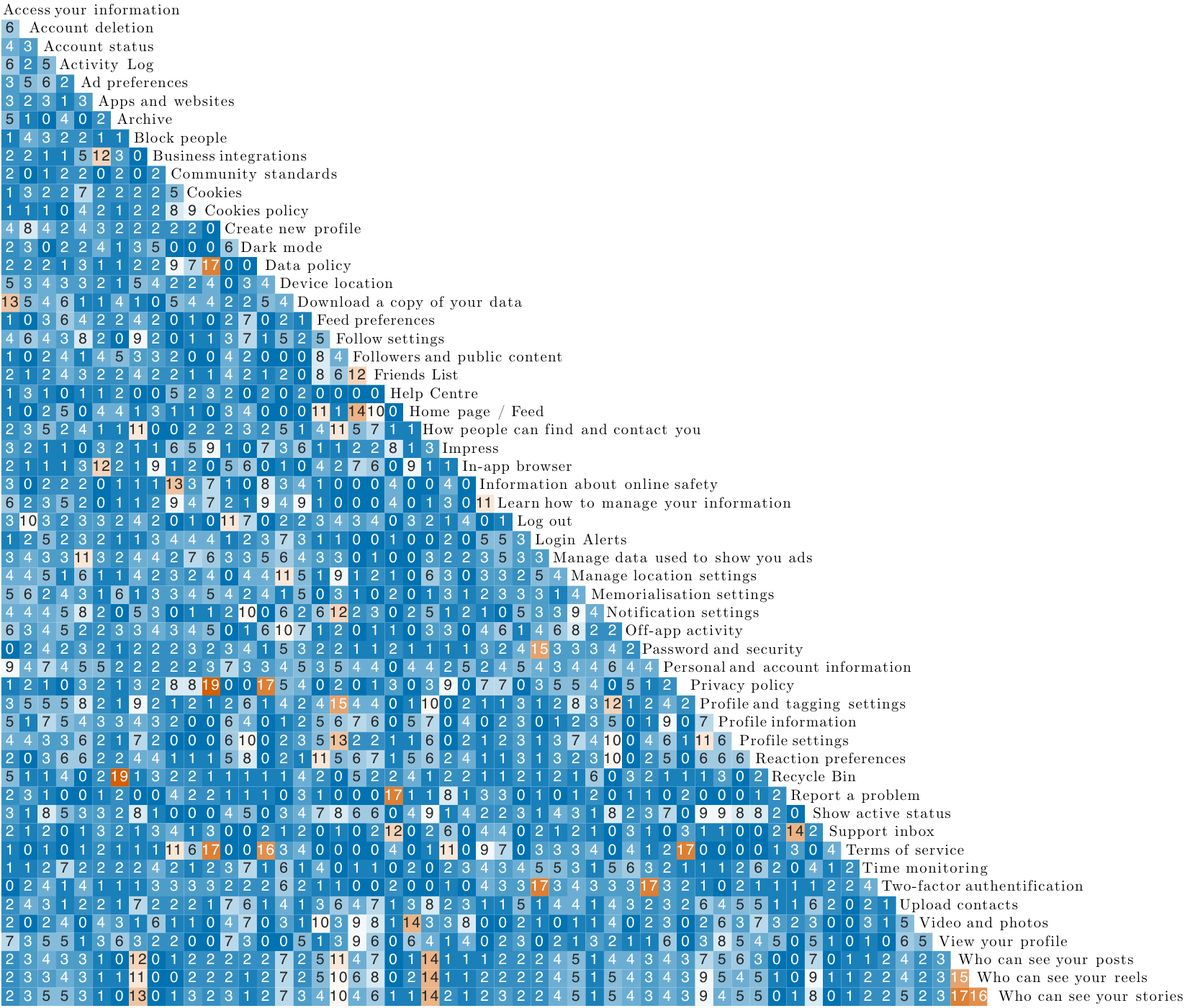}
    \caption{This figure shows the similarity matrix of all 58 Facebook features based on the 21 card sorting groups.}
    \label{fig:similarity_matrix}
    \Description{
    This figure shows the similarity matrix of all 58 cards based on the 21 card sorting groups. The similarity matrix demonstrates the relationship between any two paired cards, further utilising a heat map to enhance data comprehension.
    }
\end{figure*}

\subsection{Groups} \label{sub:groups}
The results of the card sorting task offer insights into the collective and individual perspectives of SNS users regarding the sorting of UI features. Moreover, these insights suggest common characteristics shared among SNS UI features, which, in turn, can inform an optimised structure of SNS interfaces. By enhancing the discoverability of individual features in alignment with users' expectations, interface aspects leading to labyrinthine navigation could thus be avoided.

To this end, we began our analysis by transferring the groups created by participants (see Figure~\ref{fig:cardsorting} for an example) into a similarity matrix, as visualised in Figure~\ref{fig:similarity_matrix}. This allows us to assess how often participants paired UI features, giving us a first impression of how UI features could be structured. Here, we report noteworthy similarity pairs in percentages based on the number of times the 21 participants paired up individual features. Often used features of SNS were frequently coupled together, such as `Home feed' and `Followers and public content' ($66.7\%$). Furthermore, policies were often paired (i.e. `Data policy' and `Cookies policy' at $80.9\%$).



\begin{figure*}[t!]
    \centering
    \includegraphics[width=\textwidth]{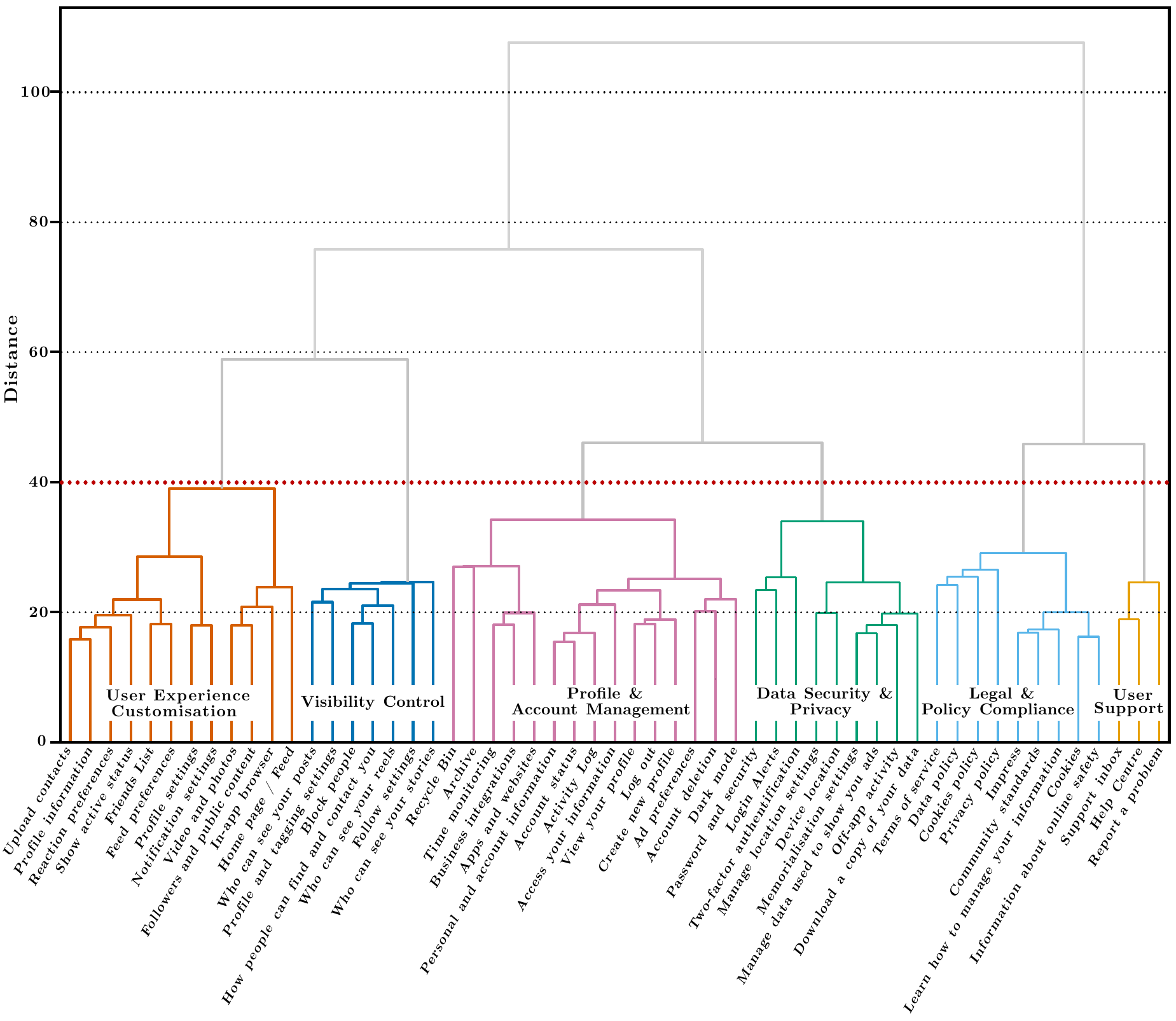}
    \caption{This hierarchical clustering dendrogram illustrates the optimal number of groups of the card sorting study, highlighted by the dotted red line at 40\%.}
    \Description{
    This hierarchical clustering dendrogram illustrates the optimal number of groups in the card sorting study. A marker highlights the 40 per cent cut-off, resulting in six groups.
    }
    \label{fig:dendrogram}
\end{figure*}

Using the data from the similarity matrix, we proceeded with a hierarchical clustering approach to determine groups with the highest agreement across participants. 
Figure~\ref{fig:dendrogram} illustrates the resulting groups. We followed common practice for the hierarchical clustering of our data by using the linkage criterion `ward' and Euclidean distance. We visually inspected the dendrogram and discussed different cut-off distances to identify meaningful groups among the authors of this work. We chose a cut-off at an Euclidean distance of 40, resulting in six groups (indicated in Figure~\ref{fig:dendrogram}), containing between 3 and 15 features each, with an average of 9 features per group. These results echo similar findings of a related approach by Nawaz~\cite{nawaz_comparison_2012}. Figure~\ref{fig:dendrogram} visualises a complete overview of the six groups, their sizes, and their corresponding features. 

\subsection{Measuring Importance}
Alongside our investigation into how participants organised UI features, we also considered their evaluations of the importance of individual features and the frequency with which they use them. These results can inform interface designs to consider perceived importance as a criterion when structuring SNS UIs.
We evaluated the data based on the ratings users' gave each card. We used outlier detection and descriptive statistics to identify significantly important or frequently used features. Here, we report the most interesting findings, while we include the full data in the supplementary material of this paper.

Using a Z-score analysis to identify outliers (with a threshold of $Z > 2$), we found two features to be significantly unimportant compared to other UI features:  `Impress' ($Z=-2.08$ and `Upload contacts' ($Z=-2.36$). Based on our analysis, there were no significantly important features.
The average importance rating for all 58 cards was relatively high, with a mean of 3.34 ($sd=0.74$). Interestingly, the highest ratings were given to `Password and Security' (mean = 4.70, $sd =0.56$), `Account Deletion' (mean = 4.45 $sd = 0.97$), and `Home Page / Feed' (mean = 4.45, $sd = 0.74$). On the other end, the UI features with the lowest importance ratings were `Show Active Status' (mean = 2.10, $sd = 0.99$), `Impress' ( mean = 1.80, sd = 0.93), and `Upload Contacts' (mean = 1.60, sd =  1.07) as the least important feature.


\subsection{Measuring Frequency}
To better understand how relevant participants perceive individual UI features, we asked them to change a card's colour (see Figure~\ref{fig:cardsorting} for an example) depending on the frequency with which they use it. In this regard, \textbf{\textcolor{cgreen}{green}} means the feature is often used, \textbf{\textcolor{cyellow}{yellow}} means the feature is moderately used, and \textbf{\textcolor{cred}{red}} means the feature is rarely used. For our analysis, we mapped the three colours to values from 1 to 3 (1=\textbf{\textcolor{cred}{red}}, 2=\textbf{\textcolor{cyellow}{yellow}}, 3= \textbf{\textcolor{cgreen}{green}}) in the form of a 3-point Likert-scale~\cite{jacoby_1971_three}. Similar to the importance ratings, we focus on important findings while we include the complete data in the supplementary material of this paper.
We used a Z-score analysis to identify outliers (with a threshold of $Z > 2$). We found two features to be significantly more often used: `Video and photos' ($Z=2.16$) and `Home page / Feed' ($Z=2.94$). Based on our analysis, there were no significantly rarely used features.

Following descriptive statistics, all 58 cards collectively featured an average score of 1.67 ($sd=0.45$) regarding usage frequency. The most frequently used UI features were `Home Page / Feed' (mean = 3.00, $sd = 0,00$), `Video and Photos' (mean = 2.65, $sd = 0.57$), and `View Your Profile' (mean = 2.40, $sd = 0.66$). In terms of least frequently used features, we find `Memorialisation Settings' (mean = 1.05, $sd = 0.22$), `Community Standards' (mean = 1.00, $sd = 0.00$), and `Terms of Service' (mean = 1.00, $sd = 0.00$). Notably, three of these UI features were rated with a standard deviation of 0.00, suggesting 100\% agreement between participants. To further investigate the results of the card sorting task, we continued by assessing the overall agreement between ratings.





\subsection{Agreement}
The agreement between participants across UI features in our data indicates similar user expectations regarding how important they find certain UI features and how frequently they use them. To investigate these notions further, we computed the percentage agreement for each feature and across all ratings in terms of importance and frequency. 
Across all importance scores, we find an average agreement of $44\%$. Furthermore, the features `Password and Security' ($75.0\%$), `Upload Contacts' ($70.0\%$), and `Account Deletion' ($70.0\%$) have the highest agreement among participants. On the other hand, the features `Ad Preferences' ($30.0\%$), `Help Centre' ($30.0\%$), and `Privacy Policy' ($25.0\%$) have the lowest agreement.
For the frequency scores, we noticed an average agreement of $62\%$. Furthermore, we find that the UI features `Community Standards', `Home page / Feed', and `Terms of service' have an agreement of $100\%$, while the features `In-app Browser' ($35.0\%$), `Time Monitoring' $35.0\%$, `Access Your Information' ($40.0\%$) have the lowest agreement among participants.

\section{Discussion \& Future Work}
Building on prior work~\cite{schaffner_understanding_2022,mildner_about_2023}, this research aims to identify relevant design considerations to bridge otherwise disconnected user expectations with regard to SNS UIs. Offering answers to our research question, the results of our card sorting study reveal sensible groupings of UI features and suggest a hierarchical structuring to afford user expectations.
In our discussion, we begin by reiterating how SNS users' expectations are broken in the first place. We then propose design considerations concerning the structure of individual UI features based on the ratings given by our participants. Afterwards, we continue with design considerations focusing on the general grouping of SNS UI features based on our hierarchical analysis.

\subsection{Aligning Expectations of SNS Users}
It is worth mentioning that we cannot know whether certain UI strategies are deployed with malicious intent; however, users face negative consequences, for instance, in the form of compromised usability and difficult-to-maintain privacy settings. 
These negative consequences are documented by a series of related work~\cite{gunawan_comparative_2021,lyngs_2020,lukoff_how_2021,mildner_ethical_2021}, including a study conducted by Schaffner et al.~\cite{schaffner_understanding_2022}, who identified various unethical practices throughout users' attempts to delete their accounts -- ultimately limiting their agency over their own account and data. 
In contrast to decreased agency and studies reporting on SNS deploying dark patterns~\cite{mildner_about_2023} or on their users misusing related platforms~\cite{christakis2010internet,twenge2018,coyne2020,beyens2020effect}, SNS have the opportunity to foster social connectedness and be of great value to maintain meaningful relationships across the globe~\cite{ahn2013social,sinclair2017facebook}. It is, therefore, relevant to consider how SNS can be redesigned to offer their users a better experience. 
By letting users sort SNS UI features, we learned about the possible optimisation of features into sensible groups in terms of restructuring their UIs. Drawing from our participants' card sortings, collecting similar features in closer proximity could improve the discoverability of individual features. Especially with critical UI features, for instance, those related to maintaining personal or ad-related data, a redesign would help users to better control settings according to their preferences~\cite{mildner_ethical_2021,wang_privacy_2013} 
To this end, our findings offer guidance and reflections for countering aspects of \textit{Labyrinthine Navigation} dark patterns in today's SNS~\cite{mildner_defending_2023}.

\subsection{Considerations for Structuring SNS UI Features}
The distribution of features across the two dimensions of importance and frequency carries certain insights for the UI design of SNS --- some are easier to respect, while others require more attention. For instance, participants rated the often-used `Home Page / Feed' very important and gave it high frequency, while unpopular features like `Upload Contacts' were considered less important as they are rarely used. It would be relatively easy to meet our participants' expectations in those regards when restructuring an SNS UI. Briefly, quick access should be granted to frequently used features, while rarely used features can be nested deeper within the UI. Unfortunately, the task becomes more challenging for more complex expectations. The feature `Account Deletion', for instance, was deemed very important but is, understandably, only rarely used. The feature `In-app Browser', on the other hand, is quite frequently used but, at the same time, perceived as unimportant. Without certain care for such specific cases, it would be easy to fall back to labyrinthine interface structures that do not meet users' needs and will be difficult to navigate. 

\begin{figure*}[t!]
    \centering
    \includegraphics[width=\textwidth]{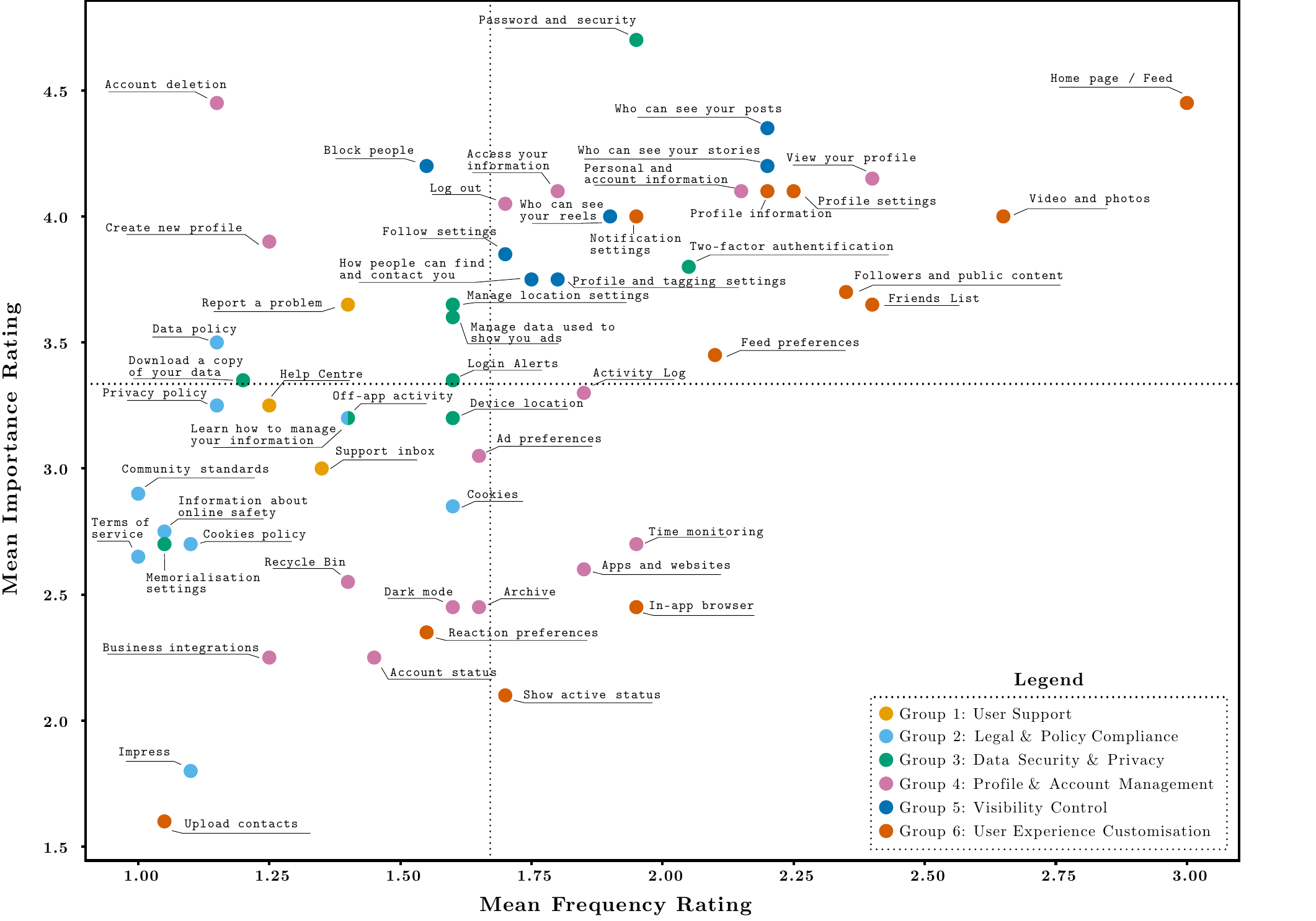}
    \caption{This figure shows a scatter plot of all 58 Facebook features based on the 21 card sorting groups placed along the two dimensions: importance and frequency. The scatter plot is divided into four quadrants through the overall mean importance and frequency ratings. The four quadrants can be characterised by containing features with low importance and low frequency ratings (lower left quadrant), low importance rating and high frequency rating (lower right quadrant), high importance and low frequency rating (upper left quadrant), and high importance and high frequency rating (upper right quadrant).}
    \Description{
    This figure shows a scatter plot of all 58 Facebook features based on the 21 card sorting groups placed along the two dimensions: importance and frequency. The scatter plot is divided into four quadrants through the overall mean importance and frequency ratings. The four quadrants can be characterised by containing features with low importance and low frequency ratings (lower left quadrant), low importance rating and high frequency rating (lower right quadrant), high importance and low frequency rating (upper left quadrant), and high importance and high frequency rating (upper right quadrant).
    }
    \label{fig:scatter_plot}
\end{figure*}

The differently perceived importance and frequency ratings of individual UI features demand good design choices to meet users' expectations, especially if a feature is perceived as important but not frequently used, or vice versa. Figure~\ref{fig:scatter_plot} shows a scatter plot of all 58 considered UI features from Facebook. Noticeably, it is divided into four quadrants through the mean importance and mean frequency ratings. The four quadrants suggest four different categories that SNS features can fall into with varying design strategies to address them: (1) High importance and high frequency ratings, (2) low importance and low frequency, (3) high importance and low frequency, (4) low importance and high frequency. For UI design, particularly the two quadrants embedding opposite high and low scores add complexity for finding sensible positions for UI features in an interface. Here, we cover each quadrant independently and propose design considerations based on traditional HCI approaches. 

\paragraph{High Importance and High Frequency}
The first quadrant is relatively straightforward. UI features that are often used and deemed important should be easy to find and engage with. Here, common practices in HCI, such as the steering-law~\cite{accot_beyond_1997}, can help UI designers find a sensible structure for these features. However, designers should be wary of overpopulating interface sections with too many options for users to choose from.

\paragraph{Low Importance and Low Frequency}
The second quadrant entails UI features that are neither frequently used nor considered important. These ratings suggest that these features do not require quick access and could be placed further into the background of the interface without breaking users' expectations. Similar to the first quadrant, this is relatively easy to address, even in combination with the first. Again, HCI principles such as the steering law can help structure the UI with respect to features within this quadrant. 

\paragraph{High Importance and Low Frequency}
These next two quadrants require more attention. UI features that SNS users find important but do not frequently need access to or use often should be positioned in the interface to allow quick access whenever needed --- even though they do not necessarily need to be omnipresent. While this may seem difficult at first, HCI has utilities to afford interactions, especially in web and app-based interfaces. UI designers could rely on interface shortcuts~\cite{bridle_shortcuts_2006} to efficiently support users' agency to access otherwise difficult-to-find features. In the same vein, searchbars~\cite{morris_searchbar_2008} allow quick and reliable access if the underlying technology can precisely interpret user input in case they are unsure of a feature's name. 

\paragraph{Low Importance and High Frequency}
Inverse to the former quadrant, UI features that are frequently used but not important to SNS users suggest potential overuse. This further implies that the features of this quadrant should not take space for other, more important features. Instead, especially if excessive or misuse is noticed, the UI structure requires a change to help users better maintain their time and regain agency of their usage behaviour~\cite{wang_privacy_2013,lukoff_how_2021}. To this end, design friction is a common design tool to help users make more reflected decisions~\cite{cox_friction_2016,leimstadtner_investigating_2023} by hindering impulsive engagement.

\subsection{Considerations for Grouping SNS UI Features}

\begin{figure*}[t!]
    \centering
    \includegraphics[width=\textwidth]{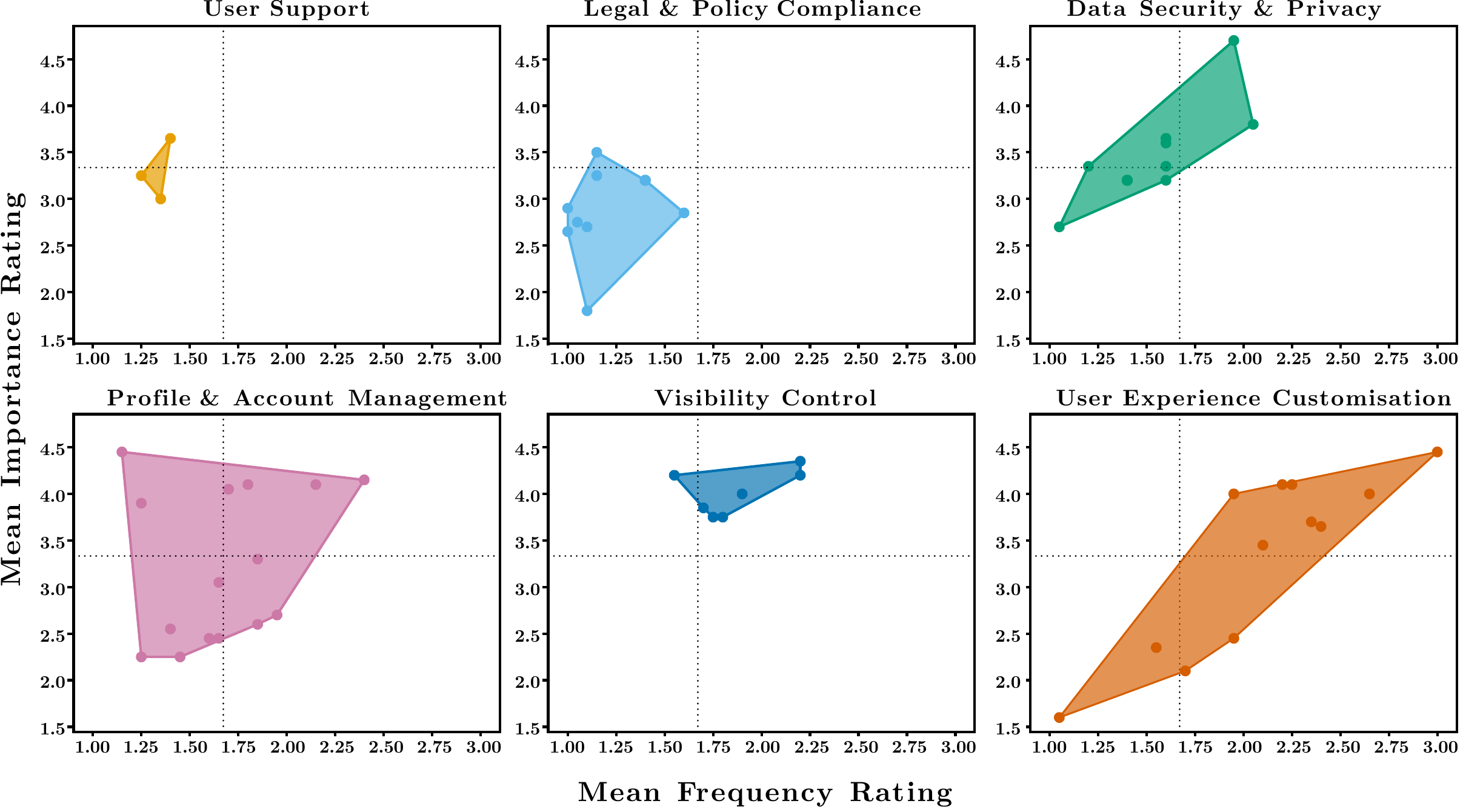}
    \caption{This figure shows each feature and their convex hull for each of the six groups as introduced in Figure~\ref{fig:scatter_plot}. Each sub-figure visualises the distribution of the contained group across the two dimensions of importance and frequency.}
    \Description{With reference to the scatter plot, this figure includes six smaller versions, each conveying the UI features of a single group, including convex hulls to convey their space over the four quadrants of the scatter plot.}
    \label{fig:scatterplots_curves}
\end{figure*}

 
Thematic groups of similar UI features help to better arrange their large quantity. Although SNS already structure their UIs based on topics, the lack of discoverability suggested by related work~\cite{schaffner_understanding_2022}, as well as the \textit{Labyrinthine Navigation} dark pattern~\cite{mildner_about_2023}, implies that improvements can be made. Previously, in Section~\ref{sub:groups}, we demonstrated optimised groupings of SNS UI features based on participants' card sortings, which we further visualised in Figure~\ref{fig:dendrogram}. Here, we discuss related design considerations per UI feature group. To this end, Figure~\ref{fig:scatterplots_curves} offers an overview of six individual groups, including convex hulls, to visualise their distribution across the importance and frequency dimensions based on participants' ratings. To describe each group, we used the individual UI features they contained in order to establish overall themes that covered their general scope. In the following paragraphs, we report the contents of each of the six groups independently, as well as discuss the resulting implications in relation to the other groups. 

\paragraph{Group 1: ``User Support'' (3 Features)} 
This group contains three features for user assistance, featuring the `Help Centre', `Report a problem', and the `Support inbox' feature. Such features help users who seek solutions to various issues, ranging from technical problems to policy-related questions. The ratings for both dimensions are mid-ranged compared to other groups, with a mean importance rating of 3.30 ($sd = 1.26$) and a mean frequency rating of 1.33 ($sd = 0.54$). These ratings suggest that users do not often require access to these features and while they are not unimportant, they seem not crucial enough to be always present.

\paragraph{Group 2: ``Legal \& Policy Compliance'' (9 Features)}
Group 2 provides the legal features, including features like `Community standards', `Cookies policy', `Data policy', and `Terms of service'. These features offer access to legal documents and information. With a mean importance rating of 2.84 ($sd =1.20$) and a mean frequency rating of 1.17 ($sd =0.33$), this group has the lowest values for importance and frequency among all groups. Thus, it can be argued that these UI features of this group should be situated deeper within interfaces, not to obfuscate other, more relevant groups.

\paragraph{Group 3: ``Data Security \& Privacy'' (9 Features)}
The next group contains features like `Device location', `Download a copy of your data', `Login alerts', and `Two-factor authentication'. These UI features offer tools to users they need to secure their own data and remain informed about any attempts to compromise it. With a mean importance rating of 3.51 ($sd =1.09$) and a mean frequency rating of 1.56 (S$sd =0.65$), the ratings are higher than those for the Legal \& Policy Compliance group but lower than those for the Visibility Control group, putting it quite in the centre in terms of SNS users' expectations. 

\paragraph{Group 4: ``Profile \& Account Management'' (15 Features)} 
This group contains features connected to profile and account management and is the largest group with a total of 15 features. They provide options for `Accessing personal information', `Account deletion', and `Ad preferences', among others. The features in this group offer ways in which users can alter how they and their information appear on the platform. Featuring a mean importance rating of 3.22 ($sd =1.11$) and a mean frequency rating of 1.67 ($sd =0.69$) the group features mid-range values for both importance and frequency ratings. As a consequence of its size, this group spans features of high importance and high frequency (i.e. `View your profile') as well as low importance and low frequency (i.e. `Business integrations'). With further attention to the four quadrants between importance and frequency scores, large groups, such as this one, may require further subdivisions into smaller clusters to increase the overall discoverability of UI features.

\paragraph{Group 5: ``Visibility Control'' (7 Features)} 
Group 5 is focused on features that control online visibility, such as `Who can see your posts', `-stories', and `-reels'. These features enable users to manage their online visibility, determining how much or how little of their content is visible to different audiences. The mean importance rating for this group is 4.01 ($sd =1.03$), which is the highest among all groups, while the mean frequency rating is 1.87 ($sd =0.74$). Overall, these ratings suggest that users would expect quick and easy access to the UI features contained in this group.

\paragraph{Group 6: ``User Experience Customization'' (12 Features)}
This final group impacts how a user interacts with the platform on a daily basis. Features include `Feed preferences', `Notification settings', and controls for `Show active status' and `Upload contacts'. These options offer users a level of control over their day-to-day engagement with the platform. For this final group, the mean importance rating is 3.33 ($sd =1.04$), and the mean frequency rating is 2.10 ($sd =0.62$). Similar to the group ``Profile \& Account Management'', this sixth group contains both low importance and low frequency (i.e. `Upload contacts') as well as high importance and high frequency (i.e. `Home page / Feed') UI features. In fact, this group contains both the lowest and highest-rated features based on our study. As with the fourth group, the size of this group may require further subdivisions to address users' expectations better.\\

Generally, the overall number of UI features available to SNS users introduces obstacles hindering the discoverability of individual ones --- negatively impacting usability. Drawing from our findings, we can rethink the placement of UI features in SNS in consideration of our participants' feedback and expectations to increase the discoverability of those that participants deemed important. 
Together, Figure~\ref{fig:scatter_plot} and Figure~\ref{fig:scatterplots_curves} offer a structured overview of users' interface expectations. Moreover, the thematic groups lay some groundwork for possible future research and design directions by focusing on the role that each feature plays within SNS contexts.
Importantly, our groups are based on limited SNS UI features and a card sorting task with 21 participants. A future study may allow for more detailed insights into similar groups or even come up with additional or different groups altogether. Nonetheless, our findings offer two general implications that help to better understand how SNS interfaces should be structured to meet users' expectations.

Firstly, our six groups offer sensible themes to organise SNS features that share similar characteristics. Aligned with users' expectations, they present a first impression for restructuring and placing of existing UI features into interface panels, menus, tabs, or any other container that fits the purpose of its application. In this work, we mainly focused on the purpose of UI features and the conceptual structuring thereof. Future work should also consider individual affordances to design the features in line with users' expectations.

Secondly, the individual groups need considerations for the internal structuring of contained features according to importance and frequency ratings. While individual groups show consistency across their features, they often span multiple quadrants as contained UI features vary in importance and frequency. Moreover, the number of included UI features varies considerably, with `User Support' containing three features compared to `Profile \& Account Management', which contains fifteen. Thus, large-sized groups may require additional design consideration and further analysis, which future work could address. Throughout our analysis of Facebook's interface, we noticed specific menus listing a large number of UI features, often requiring users to scroll in order to find specific settings. In such instances, it may be interesting to recursively apply our considerations and break down clustered settings to increase the discoverability of UI features within them.

Aligning users' expectations with experience, the groups and their themes can positively inform existing usability challenges of large and often difficult-to-manoeuvre SNS interfaces. With this work, we contribute a foundation to rethink and restructure existing SNS interfaces and highlight the relevance of considering the inclusion of the importance and frequency at which UI features are visited in usability studies. In future work, we plan to build on the gained understanding to develop and study alternative SNS UI structures and their effectiveness in avoiding \textit{Labyrinthine Navigation} dark patterns. 




\section{Limitations}
As with most research, our study has several limitations that we want to disclose here and offer potential avenues for future research. In our selection of UI features, we aimed to stay agnostic to common SNS features to mitigate these limitations. However, the focus is primarily on Facebook's interface, limiting the generalisability of our findings across other SNS platforms. While Facebook can be used as a prototypical example of SNS, this focus could result in insights that are not universally applicable, given the fast-paced evolution of SNS interfaces and their differing scopes for user engagement. However, Mildner et al.~\cite{mildner_about_2023} have identified \textit{Labyrinthine Navigation} within four SNS: Facebook, Instagram, TikTok, and Twitter, hopefully mitigating this limitation to some extent. 
In our study, we developed groups and underlying themes based on the card sorting results of 21 participants of similar demographics and knowledge in HCI-related fields. While we particularly opted for these demographics to utilise their technological literacy, this introduces two limitations. First, a higher and more diverse sample could represent a wider user base with different expectations. Second, the groups are accumulated from all participants' card sorting results. Thus, the groups only reflect averaged expectations, removing individual perceptions. In this regard, we noticed further design challenges, as some SNS address particular user needs. For example, Facebook and Instagram have features specifically directed to provide businesses and professionals with useful tools to support their efforts to attract and maintain other users' engagement.
Many of these features are only important for a certain user base and, thus, often not relevant for others. 
While we focused on static interfaces, future work could investigate the feasibility and effectiveness of adaptive or customisable UIs~\cite{gajos_supple_2004} that can be tailored to meet individual user needs

\section{Conclusion}
In the last two decades, social media platforms like Facebook have become ubiquitous companions in many peoples' lives. Based on an analysis of Facebook's interface, this research provides valuable insights for rethinking and restructuring SNS interfaces in alignment with users' expectations. Through a card sorting study involving 21 participants, including considerations of the importance in which UI features are perceived and how frequently users engage with them, we identified six common interface groups allocating UI features into key SNS features that reflect users' expectations. 
In contrast to current efforts in related work, we do not introduce design interventions to labyrinthine interfaces. Instead, we provide insights based on user-centred design to restructure user interfaces of SNS from the ground up with the goal of increasing the discoverability and usability of individual SNS features.

\bibliographystyle{ACM-Reference-Format}
\bibliography{references.bib}


\begin{thebibliography}{58}


\ifx \showCODEN    \undefined \def \showCODEN     #1{\unskip}     \fi
\ifx \showDOI      \undefined \def \showDOI       #1{#1}\fi
\ifx \showISBNx    \undefined \def \showISBNx     #1{\unskip}     \fi
\ifx \showISBNxiii \undefined \def \showISBNxiii  #1{\unskip}     \fi
\ifx \showISSN     \undefined \def \showISSN      #1{\unskip}     \fi
\ifx \showLCCN     \undefined \def \showLCCN      #1{\unskip}     \fi
\ifx \shownote     \undefined \def \shownote      #1{#1}          \fi
\ifx \showarticletitle \undefined \def \showarticletitle #1{#1}   \fi
\ifx \showURL      \undefined \def \showURL       {\relax}        \fi
\providecommand\bibfield[2]{#2}
\providecommand\bibinfo[2]{#2}
\providecommand\natexlab[1]{#1}
\providecommand\showeprint[2][]{arXiv:#2}

\bibitem[\protect\citeauthoryear{??}{soc}{2020}]%
        {social_2020}
 \bibinfo{year}{2020}\natexlab{}.
\newblock \bibinfo{title}{The {Social} {Dilemma} - {A} {Netflix} {Original}
  documentary}.
\newblock
\newblock
\urldef\tempurl%
\url{https://www.thesocialdilemma.com/}
\showURL{%
\tempurl}


\bibitem[\protect\citeauthoryear{Accot and Zhai}{Accot and Zhai}{1997}]%
        {accot_beyond_1997}
\bibfield{author}{\bibinfo{person}{Johnny Accot} {and} \bibinfo{person}{Shumin
  Zhai}.} \bibinfo{year}{1997}\natexlab{}.
\newblock \showarticletitle{Beyond Fitts' Law: Models for Trajectory-Based HCI
  Tasks}. In \bibinfo{booktitle}{\emph{Proceedings of the ACM SIGCHI Conference
  on Human Factors in Computing Systems}} (Atlanta, Georgia, USA)
  \emph{(\bibinfo{series}{CHI '97})}. \bibinfo{publisher}{Association for
  Computing Machinery}, \bibinfo{address}{New York, NY, USA},
  \bibinfo{pages}{295–302}.
\newblock
\showISBNx{0897918029}
\urldef\tempurl%
\url{https://doi.org/10.1145/258549.258760}
\showDOI{\tempurl}


\bibitem[\protect\citeauthoryear{ACM}{ACM}{2023}]%
        {acm_words_2023}
\bibfield{author}{\bibinfo{person}{ACM}.} \bibinfo{year}{2023}\natexlab{}.
\newblock \bibinfo{title}{Words matter: {Alternatives} for charged terminology
  in the computing profession}.
\newblock
\newblock
\urldef\tempurl%
\url{https://www.acm.org/diversity-inclusion/words-matter}
\showURL{%
\tempurl}


\bibitem[\protect\citeauthoryear{Ahn and Shin}{Ahn and Shin}{2013}]%
        {ahn2013social}
\bibfield{author}{\bibinfo{person}{Dohyun Ahn} {and} \bibinfo{person}{Dong-Hee
  Shin}.} \bibinfo{year}{2013}\natexlab{}.
\newblock \showarticletitle{Is the social use of media for seeking
  connectedness or for avoiding social isolation? Mechanisms underlying media
  use and subjective well-being}.
\newblock \bibinfo{journal}{\emph{Computers in Human Behavior}}
  \bibinfo{volume}{29}, \bibinfo{number}{6} (\bibinfo{year}{2013}),
  \bibinfo{pages}{2453--2462}.
\newblock


\bibitem[\protect\citeauthoryear{Alemany, Val, and García-Fornes}{Alemany
  et~al\mbox{.}}{2023}]%
        {alemany_review_2023}
\bibfield{author}{\bibinfo{person}{J. Alemany}, \bibinfo{person}{E.~Del Val},
  {and} \bibinfo{person}{A. García-Fornes}.} \bibinfo{year}{2023}\natexlab{}.
\newblock \showarticletitle{A {Review} of {Privacy} {Decision}-making
  {Mechanisms} in {Online} {Social} {Networks}}.
\newblock \bibinfo{journal}{\emph{Comput. Surveys}} \bibinfo{volume}{55},
  \bibinfo{number}{2} (\bibinfo{date}{Feb.} \bibinfo{year}{2023}),
  \bibinfo{pages}{1--32}.
\newblock
\showISSN{0360-0300, 1557-7341}
\urldef\tempurl%
\url{https://doi.org/10.1145/3494067}
\showDOI{\tempurl}


\bibitem[\protect\citeauthoryear{Alexandrovsky, Friehs, Grittner, Putze, Birk,
  Malaka, and Mandryk}{Alexandrovsky et~al\mbox{.}}{2021}]%
        {alexandrovsky_serious_2021}
\bibfield{author}{\bibinfo{person}{Dmitry Alexandrovsky},
  \bibinfo{person}{Maximilian~Achim Friehs}, \bibinfo{person}{Jendrik
  Grittner}, \bibinfo{person}{Susanne Putze}, \bibinfo{person}{Max~V. Birk},
  \bibinfo{person}{Rainer Malaka}, {and} \bibinfo{person}{Regan~L Mandryk}.}
  \bibinfo{year}{2021}\natexlab{}.
\newblock \showarticletitle{Serious {Snacking}: {A} {Survival} {Analysis} of
  how {Snacking} {Mechanics} {Affect} {Attrition} in a {Mobile} {Serious}
  {Game}}.
\newblock In \bibinfo{booktitle}{\emph{Proceedings of the 2021 {CHI}
  {Conference} on {Human} {Factors} in {Computing} {Systems}}}. Number 113.
  \bibinfo{publisher}{Association for Computing Machinery},
  \bibinfo{address}{New York, NY, USA}, \bibinfo{pages}{1--18}.
\newblock
\showISBNx{978-1-4503-8096-6}
\urldef\tempurl%
\url{https://doi.org/10.1145/3411764.3445689}
\showURL{%
\tempurl}


\bibitem[\protect\citeauthoryear{Andalibi and Buss}{Andalibi and Buss}{2020}]%
        {andalibi_human_2020}
\bibfield{author}{\bibinfo{person}{Nazanin Andalibi} {and}
  \bibinfo{person}{Justin Buss}.} \bibinfo{year}{2020}\natexlab{}.
\newblock \showarticletitle{The {Human} in {Emotion} {Recognition} on {Social}
  {Media}: {Attitudes}, {Outcomes}, {Risks}}. In
  \bibinfo{booktitle}{\emph{Proceedings of the 2020 {CHI} {Conference} on
  {Human} {Factors} in {Computing} {Systems}}}. \bibinfo{publisher}{ACM},
  \bibinfo{address}{New York, NY, USA}, \bibinfo{pages}{1--16}.
\newblock
\showISBNx{978-1-4503-6708-0}
\urldef\tempurl%
\url{https://doi.org/10.1145/3313831.3376680}
\showDOI{\tempurl}


\bibitem[\protect\citeauthoryear{Bennett, Metatla, Roudaut, and Mekler}{Bennett
  et~al\mbox{.}}{2023}]%
        {bennett_how_2023}
\bibfield{author}{\bibinfo{person}{Dan Bennett}, \bibinfo{person}{Oussama
  Metatla}, \bibinfo{person}{Anne Roudaut}, {and} \bibinfo{person}{Elisa~D.
  Mekler}.} \bibinfo{year}{2023}\natexlab{}.
\newblock \showarticletitle{How does {HCI} {Understand} {Human} {Agency} and
  {Autonomy}?}. In \bibinfo{booktitle}{\emph{Proceedings of the 2023 {CHI}
  {Conference} on {Human} {Factors} in {Computing} {Systems}}}.
  \bibinfo{publisher}{ACM}, \bibinfo{address}{Hamburg Germany},
  \bibinfo{pages}{1--18}.
\newblock
\showISBNx{978-1-4503-9421-5}
\urldef\tempurl%
\url{https://doi.org/10.1145/3544548.3580651}
\showDOI{\tempurl}


\bibitem[\protect\citeauthoryear{Beyens, Pouwels, van Driel, Keijsers, and
  Valkenburg}{Beyens et~al\mbox{.}}{2020}]%
        {beyens2020effect}
\bibfield{author}{\bibinfo{person}{Ine Beyens}, \bibinfo{person}{J~Loes
  Pouwels}, \bibinfo{person}{Irene~I van Driel}, \bibinfo{person}{Loes
  Keijsers}, {and} \bibinfo{person}{Patti~M Valkenburg}.}
  \bibinfo{year}{2020}\natexlab{}.
\newblock \showarticletitle{The effect of social media on well-being differs
  from adolescent to adolescent}.
\newblock \bibinfo{journal}{\emph{Scientific Reports}} \bibinfo{volume}{10},
  \bibinfo{number}{1} (\bibinfo{year}{2020}), \bibinfo{pages}{1--11}.
\newblock


\bibitem[\protect\citeauthoryear{Bongard-Blanchy, Rossi, Rivas, Doublet,
  Koenig, and Lenzini}{Bongard-Blanchy et~al\mbox{.}}{2021}]%
        {bongard-blanchy_i_2021}
\bibfield{author}{\bibinfo{person}{Kerstin Bongard-Blanchy},
  \bibinfo{person}{Arianna Rossi}, \bibinfo{person}{Salvador Rivas},
  \bibinfo{person}{Sophie Doublet}, \bibinfo{person}{Vincent Koenig}, {and}
  \bibinfo{person}{Gabriele Lenzini}.} \bibinfo{year}{2021}\natexlab{}.
\newblock \showarticletitle{”I Am Definitely Manipulated, Even When I Am
  Aware of It. It’s Ridiculous!” - Dark Patterns from the End-User
  Perspective}. In \bibinfo{booktitle}{\emph{Proceedings of the 2021 ACM
  Designing Interactive Systems Conference}} (Virtual Event, USA)
  \emph{(\bibinfo{series}{DIS '21})}. \bibinfo{publisher}{Association for
  Computing Machinery}, \bibinfo{address}{New York, NY, USA},
  \bibinfo{pages}{763–776}.
\newblock
\showISBNx{9781450384766}
\urldef\tempurl%
\url{https://doi.org/10.1145/3461778.3462086}
\showDOI{\tempurl}


\bibitem[\protect\citeauthoryear{Bridle and McCreath}{Bridle and
  McCreath}{2006}]%
        {bridle_shortcuts_2006}
\bibfield{author}{\bibinfo{person}{Robert Bridle} {and} \bibinfo{person}{Eric
  McCreath}.} \bibinfo{year}{2006}\natexlab{}.
\newblock \showarticletitle{Inducing Shortcuts on a Mobile Phone Interface}. In
  \bibinfo{booktitle}{\emph{Proceedings of the 11th International Conference on
  Intelligent User Interfaces}} (Sydney, Australia) \emph{(\bibinfo{series}{IUI
  '06})}. \bibinfo{publisher}{Association for Computing Machinery},
  \bibinfo{address}{New York, NY, USA}, \bibinfo{pages}{327–329}.
\newblock
\showISBNx{1595932879}
\urldef\tempurl%
\url{https://doi.org/10.1145/1111449.1111526}
\showDOI{\tempurl}


\bibitem[\protect\citeauthoryear{Brignull}{Brignull}{2023}]%
        {brignull_deceptive_2023}
\bibfield{author}{\bibinfo{person}{Harry Brignull}.}
  \bibinfo{year}{2023}\natexlab{}.
\newblock \bibinfo{title}{Deceptive {Patterns} - {Types} of {Deceptive}
  {Pattern}}.
\newblock
\newblock
\urldef\tempurl%
\url{https://www.deceptive.design/types}
\showURL{%
\tempurl}


\bibitem[\protect\citeauthoryear{Capra}{Capra}{2005}]%
        {capra_factor_2005}
\bibfield{author}{\bibinfo{person}{Miranda~G Capra}.}
  \bibinfo{year}{2005}\natexlab{}.
\newblock \showarticletitle{Factor {Analysis} of {Card} {Sort} {Data}: {An}
  {Alternative} to {Hierarchical} {Cluster} {Analysis}}.
\newblock \bibinfo{journal}{\emph{Proceedings of the Human Factors and
  Ergonomics Society Annual Meeting}} \bibinfo{volume}{49}, \bibinfo{number}{5}
  (\bibinfo{year}{2005}), \bibinfo{pages}{691--695}.
\newblock
\urldef\tempurl%
\url{https://doi.org/10.1177/154193120504900512}
\showDOI{\tempurl}


\bibitem[\protect\citeauthoryear{Chen, Piao, Lan, Cao, Gao, Lu, and Li}{Chen
  et~al\mbox{.}}{2022}]%
        {chen_practitioners_2022}
\bibfield{author}{\bibinfo{person}{Zhilong Chen}, \bibinfo{person}{Jinghua
  Piao}, \bibinfo{person}{Xiaochong Lan}, \bibinfo{person}{Hancheng Cao},
  \bibinfo{person}{Chen Gao}, \bibinfo{person}{Zhicong Lu}, {and}
  \bibinfo{person}{Yong Li}.} \bibinfo{year}{2022}\natexlab{}.
\newblock \showarticletitle{Practitioners Versus Users: A Value-Sensitive
  Evaluation of Current Industrial Recommender System Design}.
\newblock \bibinfo{journal}{\emph{Proc. ACM Hum.-Comput. Interact.}}
  \bibinfo{volume}{6}, \bibinfo{number}{CSCW2}, Article
  \bibinfo{articleno}{533} (\bibinfo{date}{nov} \bibinfo{year}{2022}),
  \bibinfo{numpages}{32}~pages.
\newblock
\urldef\tempurl%
\url{https://doi.org/10.1145/3555646}
\showDOI{\tempurl}


\bibitem[\protect\citeauthoryear{Chivukula, Obi, Carlock, and Gray}{Chivukula
  et~al\mbox{.}}{2023}]%
        {chivukula_wrangling_2023}
\bibfield{author}{\bibinfo{person}{Shruthi~Sai Chivukula}, \bibinfo{person}{Ike
  Obi}, \bibinfo{person}{Thomas~V Carlock}, {and} \bibinfo{person}{Colin~M.
  Gray}.} \bibinfo{year}{2023}\natexlab{}.
\newblock \showarticletitle{Wrangling Ethical Design Complexity: Dilemmas,
  Tensions, and Situations}. In \bibinfo{booktitle}{\emph{Companion Publication
  of the 2023 ACM Designing Interactive Systems Conference}} (Pittsburgh, PA,
  USA) \emph{(\bibinfo{series}{DIS '23 Companion})}.
  \bibinfo{publisher}{Association for Computing Machinery},
  \bibinfo{address}{New York, NY, USA}, \bibinfo{pages}{179–183}.
\newblock
\showISBNx{9781450398985}
\urldef\tempurl%
\url{https://doi.org/10.1145/3563703.3596632}
\showDOI{\tempurl}


\bibitem[\protect\citeauthoryear{Christakis}{Christakis}{2010}]%
        {christakis2010internet}
\bibfield{author}{\bibinfo{person}{Dimitri~A Christakis}.}
  \bibinfo{year}{2010}\natexlab{}.
\newblock \showarticletitle{Internet addiction: a 21 st century epidemic?}
\newblock \bibinfo{journal}{\emph{BMC medicine}} \bibinfo{volume}{8},
  \bibinfo{number}{1} (\bibinfo{year}{2010}), \bibinfo{pages}{1--3}.
\newblock


\bibitem[\protect\citeauthoryear{Cox, Gould, Cecchinato, Iacovides, and
  Renfree}{Cox et~al\mbox{.}}{2016}]%
        {cox_friction_2016}
\bibfield{author}{\bibinfo{person}{Anna~L. Cox}, \bibinfo{person}{Sandy~J.J.
  Gould}, \bibinfo{person}{Marta~E. Cecchinato}, \bibinfo{person}{Ioanna
  Iacovides}, {and} \bibinfo{person}{Ian Renfree}.}
  \bibinfo{year}{2016}\natexlab{}.
\newblock \showarticletitle{Design Frictions for Mindful Interactions: The Case
  for Microboundaries}. In \bibinfo{booktitle}{\emph{Proceedings of the 2016
  CHI Conference Extended Abstracts on Human Factors in Computing Systems}}
  (San Jose, California, USA) \emph{(\bibinfo{series}{CHI EA '16})}.
  \bibinfo{publisher}{Association for Computing Machinery},
  \bibinfo{address}{New York, NY, USA}, \bibinfo{pages}{1389–1397}.
\newblock
\showISBNx{9781450340823}
\urldef\tempurl%
\url{https://doi.org/10.1145/2851581.2892410}
\showDOI{\tempurl}


\bibitem[\protect\citeauthoryear{Coyne, Rogers, Zurcher, Stockdale, and
  Booth}{Coyne et~al\mbox{.}}{2020}]%
        {coyne2020}
\bibfield{author}{\bibinfo{person}{Sarah~M. Coyne}, \bibinfo{person}{Adam~A.
  Rogers}, \bibinfo{person}{Jessica~D. Zurcher}, \bibinfo{person}{Laura
  Stockdale}, {and} \bibinfo{person}{McCall Booth}.}
  \bibinfo{year}{2020}\natexlab{}.
\newblock \showarticletitle{Does time spent using social media impact mental
  health?: An eight year longitudinal study}.
\newblock \bibinfo{journal}{\emph{Computers in Human Behavior}}
  \bibinfo{volume}{104} (\bibinfo{year}{2020}), \bibinfo{pages}{106160}.
\newblock
\showISSN{0747-5632}
\urldef\tempurl%
\url{https://doi.org/10.1016/j.chb.2019.106160}
\showDOI{\tempurl}


\bibitem[\protect\citeauthoryear{Dhingra and Mudgal}{Dhingra and
  Mudgal}{2019}]%
        {dhingra_historical_2019}
\bibfield{author}{\bibinfo{person}{Manish Dhingra} {and}
  \bibinfo{person}{Rakesh~K. Mudgal}.} \bibinfo{year}{2019}\natexlab{}.
\newblock \bibinfo{title}{Historical {Evolution} of {Social} {Media}: {An}
  {Overview}}.
\newblock
\newblock
\urldef\tempurl%
\url{https://doi.org/10.2139/ssrn.3395665}
\showDOI{\tempurl}


\bibitem[\protect\citeauthoryear{Di~Geronimo, Braz, Fregnan, Palomba, and
  Bacchelli}{Di~Geronimo et~al\mbox{.}}{2020}]%
        {digeronimo2020}
\bibfield{author}{\bibinfo{person}{Linda Di~Geronimo}, \bibinfo{person}{Larissa
  Braz}, \bibinfo{person}{Enrico Fregnan}, \bibinfo{person}{Fabio Palomba},
  {and} \bibinfo{person}{Alberto Bacchelli}.} \bibinfo{year}{2020}\natexlab{}.
\newblock \showarticletitle{UI Dark Patterns and Where to Find Them: A Study on
  Mobile Applications and User Perception}. In
  \bibinfo{booktitle}{\emph{Proceedings of the 2020 CHI Conference on Human
  Factors in Computing Systems}} (Honolulu, HI, USA)
  \emph{(\bibinfo{series}{CHI '20})}. \bibinfo{publisher}{Association for
  Computing Machinery}, \bibinfo{address}{New York, NY, USA},
  \bibinfo{pages}{1–14}.
\newblock
\showISBNx{9781450367080}
\urldef\tempurl%
\url{https://doi.org/10.1145/3313831.3376600}
\showDOI{\tempurl}


\bibitem[\protect\citeauthoryear{Eachempati, Muzellec, and Jha}{Eachempati
  et~al\mbox{.}}{2022}]%
        {prajwal_examining_2022}
\bibfield{author}{\bibinfo{person}{Prajwal Eachempati},
  \bibinfo{person}{Laurent Muzellec}, {and} \bibinfo{person}{Ashish~Kumar
  Jha}.} \bibinfo{year}{2022}\natexlab{}.
\newblock \showarticletitle{Examining the Relationship between Privacy Setting
  Policy, Public Discourse, Business Models and Financial Performance of
  Facebook (2004–2021)}. In \bibinfo{booktitle}{\emph{Proceedings of the
  Central and Eastern European EDem and EGov Days}} (Budapest, Hungary)
  \emph{(\bibinfo{series}{CEEeGov '22})}. \bibinfo{publisher}{Association for
  Computing Machinery}, \bibinfo{address}{New York, NY, USA},
  \bibinfo{pages}{159–168}.
\newblock
\showISBNx{9781450397667}
\urldef\tempurl%
\url{https://doi.org/10.1145/3551504.3551557}
\showDOI{\tempurl}


\bibitem[\protect\citeauthoryear{Ernala, Burke, Leavitt, and Ellison}{Ernala
  et~al\mbox{.}}{2020}]%
        {ernala_how_2020}
\bibfield{author}{\bibinfo{person}{Sindhu~Kiranmai Ernala},
  \bibinfo{person}{Moira Burke}, \bibinfo{person}{Alex Leavitt}, {and}
  \bibinfo{person}{Nicole~B. Ellison}.} \bibinfo{year}{2020}\natexlab{}.
\newblock \showarticletitle{How {Well} {Do} {People} {Report} {Time} {Spent} on
  {Facebook}? {An} {Evaluation} of {Established} {Survey} {Questions} with
  {Recommendations}}.
\newblock In \bibinfo{booktitle}{\emph{Proceedings of the 2020 {CHI}
  {Conference} on {Human} {Factors} in {Computing} {Systems}}}.
  \bibinfo{publisher}{Association for Computing Machinery},
  \bibinfo{address}{New York, NY, USA}, \bibinfo{pages}{1--14}.
\newblock
\showISBNx{978-1-4503-6708-0}
\urldef\tempurl%
\url{https://doi.org/10.1145/3313831.3376435}
\showURL{%
\tempurl}


\bibitem[\protect\citeauthoryear{Fogg}{Fogg}{2009}]%
        {fogg_persuasive_2009}
\bibfield{author}{\bibinfo{person}{BJ Fogg}.} \bibinfo{year}{2009}\natexlab{}.
\newblock \showarticletitle{A Behavior Model for Persuasive Design}. In
  \bibinfo{booktitle}{\emph{Proceedings of the 4th International Conference on
  Persuasive Technology}} (Claremont, California, USA)
  \emph{(\bibinfo{series}{Persuasive '09})}. \bibinfo{publisher}{Association
  for Computing Machinery}, \bibinfo{address}{New York, NY, USA}, Article
  \bibinfo{articleno}{40}, \bibinfo{numpages}{7}~pages.
\newblock
\showISBNx{9781605583761}
\urldef\tempurl%
\url{https://doi.org/10.1145/1541948.1541999}
\showDOI{\tempurl}


\bibitem[\protect\citeauthoryear{Friedman, Kahn, Borning, and
  Huldtgren}{Friedman et~al\mbox{.}}{2013}]%
        {doorn_value_2013}
\bibfield{author}{\bibinfo{person}{Batya Friedman}, \bibinfo{person}{Peter~H.
  Kahn}, \bibinfo{person}{Alan Borning}, {and} \bibinfo{person}{Alina
  Huldtgren}.} \bibinfo{year}{2013}\natexlab{}.
\newblock \showarticletitle{Value {Sensitive} {Design} and {Information}
  {Systems}}.
\newblock In \bibinfo{booktitle}{\emph{Early engagement and new technologies:
  {Opening} up the laboratory}}, \bibfield{editor}{\bibinfo{person}{Neelke
  Doorn}, \bibinfo{person}{Daan Schuurbiers}, \bibinfo{person}{Ibo van~de
  Poel}, {and} \bibinfo{person}{Michael~E. Gorman}} (Eds.).
  Vol.~\bibinfo{volume}{16}. \bibinfo{publisher}{Springer Netherlands},
  \bibinfo{address}{Dordrecht}, \bibinfo{pages}{55--95}.
\newblock
\showISBNx{978-94-007-7843-6 978-94-007-7844-3}
\urldef\tempurl%
\url{https://doi.org/10.1007/978-94-007-7844-3_4}
\showDOI{\tempurl}
\newblock
\shownote{Series Title: Philosophy of Engineering and Technology.}


\bibitem[\protect\citeauthoryear{Gajos and Weld}{Gajos and Weld}{2004}]%
        {gajos_supple_2004}
\bibfield{author}{\bibinfo{person}{Krzysztof Gajos} {and}
  \bibinfo{person}{Daniel~S. Weld}.} \bibinfo{year}{2004}\natexlab{}.
\newblock \showarticletitle{SUPPLE: automatically generating user interfaces}.
  In \bibinfo{booktitle}{\emph{Proceedings of the 9th International Conference
  on Intelligent User Interfaces}} (Funchal, Madeira, Portugal)
  \emph{(\bibinfo{series}{IUI '04})}. \bibinfo{publisher}{Association for
  Computing Machinery}, \bibinfo{address}{New York, NY, USA},
  \bibinfo{pages}{93–100}.
\newblock
\showISBNx{1581138156}
\urldef\tempurl%
\url{https://doi.org/10.1145/964442.964461}
\showDOI{\tempurl}


\bibitem[\protect\citeauthoryear{Gray, Bielova, Santos, and Mildner}{Gray
  et~al\mbox{.}}{2024}]%
        {gray_ontology_2024}
\bibfield{author}{\bibinfo{person}{Colin~M Gray}, \bibinfo{person}{Nataliia
  Bielova}, \bibinfo{person}{Cristiana Santos}, {and} \bibinfo{person}{Thomas
  Mildner}.} \bibinfo{year}{2024}\natexlab{}.
\newblock \showarticletitle{An Ontology of Dark Patterns: Foundations,
  Definitions, and a Structure for Transdisciplinary Action}.
\newblock \bibinfo{journal}{\emph{arXiv preprint arXiv:2309.09640}}
  (\bibinfo{year}{2024}).
\newblock


\bibitem[\protect\citeauthoryear{Gray and Chivukula}{Gray and
  Chivukula}{2019}]%
        {gray_ethical_2019}
\bibfield{author}{\bibinfo{person}{Colin~M. Gray} {and}
  \bibinfo{person}{Shruthi~Sai Chivukula}.} \bibinfo{year}{2019}\natexlab{}.
\newblock \showarticletitle{Ethical {Mediation} in {UX} {Practice}}.
\newblock In \bibinfo{booktitle}{\emph{Proceedings of the 2019 {CHI}
  {Conference} on {Human} {Factors} in {Computing} {Systems}}}.
  \bibinfo{publisher}{Association for Computing Machinery},
  \bibinfo{address}{New York, NY, USA}, \bibinfo{pages}{1--11}.
\newblock
\showISBNx{978-1-4503-5970-2}
\urldef\tempurl%
\url{https://doi.org/10.1145/3290605.3300408}
\showURL{%
\tempurl}


\bibitem[\protect\citeauthoryear{Gray, Kou, Battles, Hoggatt, and Toombs}{Gray
  et~al\mbox{.}}{2018}]%
        {gray_dark_2018}
\bibfield{author}{\bibinfo{person}{Colin~M. Gray}, \bibinfo{person}{Yubo Kou},
  \bibinfo{person}{Bryan Battles}, \bibinfo{person}{Joseph Hoggatt}, {and}
  \bibinfo{person}{Austin~L. Toombs}.} \bibinfo{year}{2018}\natexlab{}.
\newblock \showarticletitle{The dark (patterns) side of {UX} design}.
\newblock \bibinfo{journal}{\emph{Conference on Human Factors in Computing
  Systems - Proceedings}}  \bibinfo{volume}{2018-April} (\bibinfo{year}{2018}),
  \bibinfo{pages}{1--14}.
\newblock
\showISSN{9781450356206}
\urldef\tempurl%
\url{https://doi.org/10.1145/3173574.3174108}
\showDOI{\tempurl}


\bibitem[\protect\citeauthoryear{Gray, Santos, Bielova, and Mildner}{Gray
  et~al\mbox{.}}{[n.d.]}]%
        {gray_ontology_2023}
\bibfield{author}{\bibinfo{person}{Colin~M. Gray}, \bibinfo{person}{Cristiana
  Santos}, \bibinfo{person}{Nataliia Bielova}, {and} \bibinfo{person}{Thomas
  Mildner}.} \bibinfo{year}{[n.d.]}\natexlab{}.
\newblock \bibinfo{title}{An Ontology of Dark Patterns: Foundations,
  Definitions, and a Structure for Transdisciplinary Action}.
\newblock
\newblock
\showeprint[arxiv]{2309.09640 [cs]}
\urldef\tempurl%
\url{http://arxiv.org/abs/2309.09640}
\showURL{%
\tempurl}


\bibitem[\protect\citeauthoryear{Gunawan, Pradeep, Choffnes, Hartzog, and
  Wilson}{Gunawan et~al\mbox{.}}{2021}]%
        {gunawan_comparative_2021}
\bibfield{author}{\bibinfo{person}{Johanna Gunawan}, \bibinfo{person}{Amogh
  Pradeep}, \bibinfo{person}{David Choffnes}, \bibinfo{person}{Woodrow
  Hartzog}, {and} \bibinfo{person}{Christo Wilson}.}
  \bibinfo{year}{2021}\natexlab{}.
\newblock \showarticletitle{A {Comparative} {Study} of {Dark} {Patterns}
  {Across} {Web} and {Mobile} {Modalities}}.
\newblock \bibinfo{journal}{\emph{Proceedings of the ACM on Human-Computer
  Interaction}} \bibinfo{volume}{5}, \bibinfo{number}{CSCW2}
  (\bibinfo{date}{Oct.} \bibinfo{year}{2021}), \bibinfo{pages}{1--29}.
\newblock
\showISSN{2573-0142}
\urldef\tempurl%
\url{https://doi.org/10.1145/3479521}
\showDOI{\tempurl}


\bibitem[\protect\citeauthoryear{Habib, Pearman, Young, Saxena, Zhang, and
  Cranor}{Habib et~al\mbox{.}}{2022}]%
        {habib_identifying_2022}
\bibfield{author}{\bibinfo{person}{Hana Habib}, \bibinfo{person}{Sarah
  Pearman}, \bibinfo{person}{Ellie Young}, \bibinfo{person}{Ishika Saxena},
  \bibinfo{person}{Robert Zhang}, {and} \bibinfo{person}{Lorrie~FaIth Cranor}.}
  \bibinfo{year}{2022}\natexlab{}.
\newblock \showarticletitle{Identifying {User} {Needs} for {Advertising}
  {Controls} on {Facebook}}.
\newblock \bibinfo{journal}{\emph{Proceedings of the ACM on Human-Computer
  Interaction}} \bibinfo{volume}{6}, \bibinfo{number}{CSCW1}
  (\bibinfo{date}{March} \bibinfo{year}{2022}), \bibinfo{pages}{1--42}.
\newblock
\showISSN{2573-0142}
\urldef\tempurl%
\url{https://doi.org/10.1145/3512906}
\showDOI{\tempurl}


\bibitem[\protect\citeauthoryear{Hansen and Jespersen}{Hansen and
  Jespersen}{2013}]%
        {hansen_nudge_2013}
\bibfield{author}{\bibinfo{person}{Pelle~Guldborg Hansen} {and}
  \bibinfo{person}{Andreas~Maaløe Jespersen}.}
  \bibinfo{year}{2013}\natexlab{}.
\newblock \showarticletitle{Nudge and the {Manipulation} of {Choice}: {A}
  {Framework} for the {Responsible} {Use} of the {Nudge} {Approach} to
  {Behaviour} {Change} in {Public} {Policy}}.
\newblock \bibinfo{journal}{\emph{European Journal of Risk Regulation}}
  \bibinfo{volume}{4}, \bibinfo{number}{1} (\bibinfo{date}{March}
  \bibinfo{year}{2013}), \bibinfo{pages}{3--28}.
\newblock
\showISSN{1867-299X, 2190-8249}
\urldef\tempurl%
\url{https://doi.org/10.1017/S1867299X00002762}
\showDOI{\tempurl}


\bibitem[\protect\citeauthoryear{Jacoby and Matell}{Jacoby and Matell}{1971}]%
        {jacoby_1971_three}
\bibfield{author}{\bibinfo{person}{Jacob Jacoby} {and}
  \bibinfo{person}{Michael~S. Matell}.} \bibinfo{year}{1971}\natexlab{}.
\newblock \showarticletitle{Three-Point Likert Scales Are Good Enough}.
\newblock \bibinfo{journal}{\emph{Journal of Marketing Research}}
  \bibinfo{volume}{8}, \bibinfo{number}{4} (\bibinfo{year}{1971}),
  \bibinfo{pages}{495--500}.
\newblock
\showISSN{00222437}
\urldef\tempurl%
\url{http://www.jstor.org/stable/3150242}
\showURL{%
\tempurl}


\bibitem[\protect\citeauthoryear{Jeong, Kim, Yum, and Hwang}{Jeong
  et~al\mbox{.}}{2016}]%
        {jeong_what_2016}
\bibfield{author}{\bibinfo{person}{Se-Hoon Jeong}, \bibinfo{person}{HyoungJee
  Kim}, \bibinfo{person}{Jung-Yoon Yum}, {and} \bibinfo{person}{Yoori Hwang}.}
  \bibinfo{year}{2016}\natexlab{}.
\newblock \showarticletitle{What type of content are smartphone users addicted
  to?: {SNS} vs. games}.
\newblock \bibinfo{journal}{\emph{Computers in Human Behavior}}
  \bibinfo{volume}{54} (\bibinfo{year}{2016}), \bibinfo{pages}{10--17}.
\newblock
\showISSN{0747-5632}
\urldef\tempurl%
\url{https://doi.org/10.1016/j.chb.2015.07.035}
\showDOI{\tempurl}


\bibitem[\protect\citeauthoryear{Kokolakis}{Kokolakis}{2017}]%
        {kokolakis_privacy_2017}
\bibfield{author}{\bibinfo{person}{Spyros Kokolakis}.}
  \bibinfo{year}{2017}\natexlab{}.
\newblock \showarticletitle{Privacy attitudes and privacy behaviour: {A} review
  of current research on the privacy paradox phenomenon}.
\newblock \bibinfo{journal}{\emph{Computers \& Security}}  \bibinfo{volume}{64}
  (\bibinfo{date}{Jan.} \bibinfo{year}{2017}), \bibinfo{pages}{122--134}.
\newblock
\showISSN{01674048}
\urldef\tempurl%
\url{https://doi.org/10.1016/j.cose.2015.07.002}
\showDOI{\tempurl}


\bibitem[\protect\citeauthoryear{Leimstädtner, Sörries, and
  Müller-Birn}{Leimstädtner et~al\mbox{.}}{2023}]%
        {leimstadtner_investigating_2023}
\bibfield{author}{\bibinfo{person}{David Leimstädtner}, \bibinfo{person}{Peter
  Sörries}, {and} \bibinfo{person}{Claudia Müller-Birn}.}
  \bibinfo{year}{2023}\natexlab{}.
\newblock \showarticletitle{Investigating {Responsible} {Nudge} {Design} for
  {Informed} {Decision}-{Making} {Enabling} {Transparent} and {Reflective}
  {Decision}-{Making}}. In \bibinfo{booktitle}{\emph{Mensch und {Computer}
  2023}}. \bibinfo{publisher}{ACM}, \bibinfo{address}{Rapperswil Switzerland},
  \bibinfo{pages}{220--236}.
\newblock
\showISBNx{9798400707711}
\urldef\tempurl%
\url{https://doi.org/10.1145/3603555.3603567}
\showDOI{\tempurl}


\bibitem[\protect\citeauthoryear{Lukoff, Yu, Kientz, and Hiniker}{Lukoff
  et~al\mbox{.}}{2018}]%
        {lukoff_how_2021}
\bibfield{author}{\bibinfo{person}{Kai Lukoff}, \bibinfo{person}{Cissy Yu},
  \bibinfo{person}{Julie Kientz}, {and} \bibinfo{person}{Alexis Hiniker}.}
  \bibinfo{year}{2018}\natexlab{}.
\newblock \showarticletitle{What Makes Smartphone Use Meaningful or
  Meaningless?}, In \bibinfo{booktitle}{Proc. ACM Interact. Mob. Wearable
  Ubiquitous Technol.}
\newblock \bibinfo{journal}{\emph{Proc. ACM Interact. Mob. Wearable Ubiquitous
  Technol.}} \bibinfo{volume}{2}, \bibinfo{number}{1}, Article
  \bibinfo{articleno}{22}, \bibinfo{numpages}{26}~pages.
\newblock
\urldef\tempurl%
\url{https://doi.org/10.1145/3191754}
\showDOI{\tempurl}


\bibitem[\protect\citeauthoryear{Lyngs, Lukoff, Slovak, Seymour, Webb, Jirotka,
  Zhao, Kleek, and Shadbolt}{Lyngs et~al\mbox{.}}{2020}]%
        {lyngs_2020}
\bibfield{author}{\bibinfo{person}{Ulrik Lyngs}, \bibinfo{person}{Kai Lukoff},
  \bibinfo{person}{Petr Slovak}, \bibinfo{person}{William Seymour},
  \bibinfo{person}{Helena Webb}, \bibinfo{person}{Marina Jirotka},
  \bibinfo{person}{Jun Zhao}, \bibinfo{person}{Max~Van Kleek}, {and}
  \bibinfo{person}{Nigel Shadbolt}.} \bibinfo{year}{2020}\natexlab{}.
\newblock \showarticletitle{‘I Just want to Hack Myself to Not Get
  Distracted’: Evaluating Design Interventions for Self-Control on Facebook}.
  In \bibinfo{booktitle}{\emph{{CHI}'20}}. \bibinfo{publisher}{ACM},
  \bibinfo{address}{Honolulu}, \bibinfo{pages}{1--15}.
\newblock
\showISBNx{978-1-4503-6708-0}
\urldef\tempurl%
\url{https://doi.org/10.1145/3313831.3376672}
\showDOI{\tempurl}


\bibitem[\protect\citeauthoryear{Masaki, Shibata, Hoshino, Ishihama, Saito, and
  Yatani}{Masaki et~al\mbox{.}}{2020}]%
        {masaki_exploring_2020}
\bibfield{author}{\bibinfo{person}{Hiroaki Masaki}, \bibinfo{person}{Kengo
  Shibata}, \bibinfo{person}{Shui Hoshino}, \bibinfo{person}{Takahiro
  Ishihama}, \bibinfo{person}{Nagayuki Saito}, {and} \bibinfo{person}{Koji
  Yatani}.} \bibinfo{year}{2020}\natexlab{}.
\newblock \showarticletitle{Exploring {Nudge} {Designs} to {Help} {Adolescent}
  {SNS} {Users} {Avoid} {Privacy} and {Safety} {Threats}}. In
  \bibinfo{booktitle}{\emph{Proceedings of the 2020 {CHI} {Conference} on
  {Human} {Factors} in {Computing} {Systems}}}. \bibinfo{publisher}{ACM},
  \bibinfo{address}{Honolulu HI USA}, \bibinfo{pages}{1--11}.
\newblock
\showISBNx{978-1-4503-6708-0}
\urldef\tempurl%
\url{https://doi.org/10.1145/3313831.3376666}
\showDOI{\tempurl}


\bibitem[\protect\citeauthoryear{McIntyre}{McIntyre}{2014}]%
        {mcintyre_evolution_2014}
\bibfield{author}{\bibinfo{person}{Karen~Elizabeth McIntyre}.}
  \bibinfo{year}{2014}\natexlab{}.
\newblock \showarticletitle{The {Evolution} of {Social} {Media} from 1969 to
  2013: {A} {Change} in {Competition} and a {Trend} {Toward} {Complementary},
  {Niche} {Sites}}.
\newblock \bibinfo{journal}{\emph{The Journal of Social Media in Society}}
  \bibinfo{volume}{3}, \bibinfo{number}{2} (\bibinfo{date}{Dec.}
  \bibinfo{year}{2014}).
\newblock
\showISSN{2325-503X}
\urldef\tempurl%
\url{https://thejsms.org/index.php/JSMS/article/view/89}
\showURL{%
\tempurl}
\newblock
\shownote{Number: 2.}


\bibitem[\protect\citeauthoryear{Mildner, Freye, Savino, Doyle, Cowan, and
  Malaka}{Mildner et~al\mbox{.}}{2023a}]%
        {mildner_defending_2023}
\bibfield{author}{\bibinfo{person}{Thomas Mildner}, \bibinfo{person}{Merle
  Freye}, \bibinfo{person}{Gian-Luca Savino}, \bibinfo{person}{Philip~R.
  Doyle}, \bibinfo{person}{Benjamin~R. Cowan}, {and} \bibinfo{person}{Rainer
  Malaka}.} \bibinfo{year}{2023}\natexlab{a}.
\newblock \showarticletitle{Defending Against the Dark Arts: Recognising Dark
  Patterns in Social Media}. In \bibinfo{booktitle}{\emph{Designing Interactive
  Systems Conference (DIS '23), July 10--14, 2023, Pittsburgh, PA, USA}}
  (Pittsburgh, PA, USA) \emph{(\bibinfo{series}{DIS '23})}.
  \bibinfo{publisher}{Association for Computing Machinery},
  \bibinfo{address}{New York, NY, USA}, 13.
\newblock
\showISBNx{978-1-4503-9893-0/23/07}
\urldef\tempurl%
\url{https://doi.org/1010.1145/3563657.3595964}
\showDOI{\tempurl}


\bibitem[\protect\citeauthoryear{Mildner and Savino}{Mildner and
  Savino}{2021}]%
        {mildner_ethical_2021}
\bibfield{author}{\bibinfo{person}{Thomas Mildner} {and}
  \bibinfo{person}{Gian-Luca Savino}.} \bibinfo{year}{2021}\natexlab{}.
\newblock \showarticletitle{Ethical {User} {Interfaces}: {Exploring} the
  {Effects} of {Dark} {Patterns} on {Facebook}}. In
  \bibinfo{booktitle}{\emph{Extended {Abstracts} of the 2021 {CHI} {Conference}
  on {Human} {Factors} in {Computing} {Systems}}}. \bibinfo{publisher}{ACM},
  \bibinfo{address}{Yokohama Japan}, \bibinfo{pages}{1--7}.
\newblock
\showISBNx{978-1-4503-8095-9}
\urldef\tempurl%
\url{https://doi.org/10.1145/3411763.3451659}
\showDOI{\tempurl}


\bibitem[\protect\citeauthoryear{Mildner, Savino, Doyle, Cowan, and
  Malaka}{Mildner et~al\mbox{.}}{2023b}]%
        {mildner_about_2023}
\bibfield{author}{\bibinfo{person}{Thomas Mildner}, \bibinfo{person}{Gian-Luca
  Savino}, \bibinfo{person}{Philip~R. Doyle}, \bibinfo{person}{Benjamin~R.
  Cowan}, {and} \bibinfo{person}{Rainer Malaka}.}
  \bibinfo{year}{2023}\natexlab{b}.
\newblock \showarticletitle{About Engaging and Governing Strategies: A Thematic
  Analysis of Dark Patterns in Social Networking Services}. In
  \bibinfo{booktitle}{\emph{Proceedings of the 2023 CHI Conference on Human
  Factors in Computing Systems}} (Hamburg, Germany) \emph{(\bibinfo{series}{CHI
  '23})}. \bibinfo{publisher}{Association for Computing Machinery},
  \bibinfo{address}{New York, NY, USA}, Article \bibinfo{articleno}{192},
  \bibinfo{numpages}{15}~pages.
\newblock
\showISBNx{9781450394215}
\urldef\tempurl%
\url{https://doi.org/10.1145/3544548.3580695}
\showDOI{\tempurl}


\bibitem[\protect\citeauthoryear{Monge~Roffarello, Lukoff, and
  De~Russis}{Monge~Roffarello et~al\mbox{.}}{2023}]%
        {monge_roffarello_defining_2023}
\bibfield{author}{\bibinfo{person}{Alberto Monge~Roffarello},
  \bibinfo{person}{Kai Lukoff}, {and} \bibinfo{person}{Luigi De~Russis}.}
  \bibinfo{year}{2023}\natexlab{}.
\newblock \showarticletitle{Defining and {Identifying} {Attention} {Capture}
  {Deceptive} {Designs} in {Digital} {Interfaces}}. In
  \bibinfo{booktitle}{\emph{Proceedings of the 2023 {CHI} {Conference} on
  {Human} {Factors} in {Computing} {Systems}}}. \bibinfo{publisher}{ACM},
  \bibinfo{address}{Hamburg Germany}, \bibinfo{pages}{1--19}.
\newblock
\showISBNx{978-1-4503-9421-5}
\urldef\tempurl%
\url{https://doi.org/10.1145/3544548.3580729}
\showDOI{\tempurl}


\bibitem[\protect\citeauthoryear{Morris, Ringel~Morris, and Venolia}{Morris
  et~al\mbox{.}}{2008}]%
        {morris_searchbar_2008}
\bibfield{author}{\bibinfo{person}{Dan Morris}, \bibinfo{person}{Meredith
  Ringel~Morris}, {and} \bibinfo{person}{Gina Venolia}.}
  \bibinfo{year}{2008}\natexlab{}.
\newblock \showarticletitle{SearchBar: A Search-Centric Web History for Task
  Resumption and Information Re-Finding}. In
  \bibinfo{booktitle}{\emph{Proceedings of the SIGCHI Conference on Human
  Factors in Computing Systems}} (Florence, Italy) \emph{(\bibinfo{series}{CHI
  '08})}. \bibinfo{publisher}{Association for Computing Machinery},
  \bibinfo{address}{New York, NY, USA}, \bibinfo{pages}{1207–1216}.
\newblock
\showISBNx{9781605580111}
\urldef\tempurl%
\url{https://doi.org/10.1145/1357054.1357242}
\showDOI{\tempurl}


\bibitem[\protect\citeauthoryear{Nawaz}{Nawaz}{2012}]%
        {nawaz_comparison_2012}
\bibfield{author}{\bibinfo{person}{Ather Nawaz}.}
  \bibinfo{year}{2012}\natexlab{}.
\newblock \showarticletitle{A {Comparison} of {Card}-sorting {Analysis}
  {Methods}}. In \bibinfo{booktitle}{\emph{The 10th {Asia} {Pacific}
  {Conference} on {Computer} {Human} {Interaction}. 2012 - {Matsue}, {Japan}}},
  Vol.~\bibinfo{volume}{2}. \bibinfo{publisher}{ACM}, \bibinfo{address}{Matsue,
  Japan}, \bibinfo{pages}{583--592}.
\newblock
\showISBNx{978-4-9906562-0-1}


\bibitem[\protect\citeauthoryear{{RealtimeBoard, Inc.}}{{RealtimeBoard,
  Inc.}}{2023}]%
        {realtimeboard_inc_miro_2023}
\bibfield{author}{\bibinfo{person}{{RealtimeBoard, Inc.}}}
  \bibinfo{year}{2023}\natexlab{}.
\newblock \bibinfo{title}{Miro {\textbar} {The} {Visual} {Collaboration}
  {Platform} for {Every} {Team}}.
\newblock
\newblock
\urldef\tempurl%
\url{https://miro.com/}
\showURL{%
\tempurl}


\bibitem[\protect\citeauthoryear{Schaffner, Lingareddy, and Chetty}{Schaffner
  et~al\mbox{.}}{2022}]%
        {schaffner_understanding_2022}
\bibfield{author}{\bibinfo{person}{Brennan Schaffner}, \bibinfo{person}{Neha~A.
  Lingareddy}, {and} \bibinfo{person}{Marshini Chetty}.}
  \bibinfo{year}{2022}\natexlab{}.
\newblock \showarticletitle{Understanding {Account} {Deletion} and {Relevant}
  {Dark} {Patterns} on {Social} {Media}}.
\newblock \bibinfo{journal}{\emph{Proceedings of the ACM on Human-Computer
  Interaction}} \bibinfo{volume}{6}, \bibinfo{number}{CSCW2}
  (\bibinfo{date}{Nov.} \bibinfo{year}{2022}), \bibinfo{pages}{1--43}.
\newblock
\showISSN{2573-0142}
\urldef\tempurl%
\url{https://doi.org/10.1145/3555142}
\showDOI{\tempurl}


\bibitem[\protect\citeauthoryear{Schoenebeck}{Schoenebeck}{2014}]%
        {schoenebeck_giving_2014}
\bibfield{author}{\bibinfo{person}{Sarita~Yardi Schoenebeck}.}
  \bibinfo{year}{2014}\natexlab{}.
\newblock \showarticletitle{Giving up {Twitter} for {Lent}: how and why we take
  breaks from social media}. In \bibinfo{booktitle}{\emph{Proceedings of the
  {SIGCHI} {Conference} on {Human} {Factors} in {Computing} {Systems}}}
  \emph{(\bibinfo{series}{{CHI} '14})}. \bibinfo{publisher}{Association for
  Computing Machinery}, \bibinfo{address}{New York, NY, USA},
  \bibinfo{pages}{773--782}.
\newblock
\showISBNx{978-1-4503-2473-1}
\urldef\tempurl%
\url{https://doi.org/10.1145/2556288.2556983}
\showDOI{\tempurl}


\bibitem[\protect\citeauthoryear{Sinclair and Grieve}{Sinclair and
  Grieve}{2017}]%
        {sinclair2017facebook}
\bibfield{author}{\bibinfo{person}{Tara~J Sinclair} {and}
  \bibinfo{person}{Rachel Grieve}.} \bibinfo{year}{2017}\natexlab{}.
\newblock \showarticletitle{Facebook as a source of social connectedness in
  older adults}.
\newblock \bibinfo{journal}{\emph{Computers in Human Behavior}}
  \bibinfo{volume}{66} (\bibinfo{year}{2017}), \bibinfo{pages}{363--369}.
\newblock


\bibitem[\protect\citeauthoryear{Singer}{Singer}{2018}]%
        {singer_why_2018}
\bibfield{author}{\bibinfo{person}{Natasha Singer}.}
  \bibinfo{year}{2018}\natexlab{}.
\newblock \showarticletitle{Why the {F}.{T}.{C}. {Is} {Taking} a {New} {Look}
  at {Facebook} {Privacy}}.
\newblock \bibinfo{journal}{\emph{The New York Times}} (\bibinfo{date}{Dec.}
  \bibinfo{year}{2018}).
\newblock
\showISSN{0362-4331}
\urldef\tempurl%
\url{https://www.nytimes.com/2018/12/22/technology/facebook-consent-decree-details.html}
\showURL{%
\tempurl}


\bibitem[\protect\citeauthoryear{Stone, Guan, LaBarbera, Ceren, Garcia, Huie,
  Stump, and Wang}{Stone et~al\mbox{.}}{2022}]%
        {stone_why_2022}
\bibfield{author}{\bibinfo{person}{Charles~B. Stone}, \bibinfo{person}{Li
  Guan}, \bibinfo{person}{Gabriella LaBarbera}, \bibinfo{person}{Melissa
  Ceren}, \bibinfo{person}{Brandon Garcia}, \bibinfo{person}{Kelly Huie},
  \bibinfo{person}{Carissa Stump}, {and} \bibinfo{person}{Qi Wang}.}
  \bibinfo{year}{2022}\natexlab{}.
\newblock \showarticletitle{Why do people share memories online? {An}
  examination of the motives and characteristics of social media users}.
\newblock \bibinfo{journal}{\emph{Memory}} \bibinfo{volume}{30},
  \bibinfo{number}{4} (\bibinfo{year}{2022}), \bibinfo{pages}{450--464}.
\newblock
\urldef\tempurl%
\url{https://doi.org/10.1080/09658211.2022.2040534}
\showDOI{\tempurl}
\newblock
\shownote{Publisher: Routledge \_eprint:
  https://doi.org/10.1080/09658211.2022.2040534.}


\bibitem[\protect\citeauthoryear{Thaler and Sunstein}{Thaler and
  Sunstein}{2008}]%
        {thaler_nudge_2008}
\bibfield{author}{\bibinfo{person}{Richard~H. Thaler} {and}
  \bibinfo{person}{Cass~R. Sunstein}.} \bibinfo{year}{2008}\natexlab{}.
\newblock \bibinfo{booktitle}{\emph{Nudge: improving decisions about health,
  wealth, and happiness}}.
\newblock \bibinfo{publisher}{Yale University Press}, \bibinfo{address}{New
  Haven}.
\newblock
\showISBNx{978-0-300-12223-7}
\newblock
\shownote{OCLC: ocn181517463.}


\bibitem[\protect\citeauthoryear{Twenge, Joiner, Rogers, and Martin}{Twenge
  et~al\mbox{.}}{2018}]%
        {twenge2018}
\bibfield{author}{\bibinfo{person}{Jean~M. Twenge}, \bibinfo{person}{Thomas~E.
  Joiner}, \bibinfo{person}{Megan~L. Rogers}, {and}
  \bibinfo{person}{Gabrielle~N. Martin}.} \bibinfo{year}{2018}\natexlab{}.
\newblock \showarticletitle{Increases in Depressive Symptoms, Suicide-Related
  Outcomes, and Suicide Rates Among U.S. Adolescents After 2010 and Links to
  Increased New Media Screen Time}.
\newblock \bibinfo{journal}{\emph{Clinical Psychological Science}}
  \bibinfo{volume}{6}, \bibinfo{number}{1} (\bibinfo{year}{2018}),
  \bibinfo{pages}{3--17}.
\newblock
\urldef\tempurl%
\url{https://doi.org/10.1177/2167702617723376}
\showDOI{\tempurl}


\bibitem[\protect\citeauthoryear{Wang, Jackson, Gaskin, and Wang}{Wang
  et~al\mbox{.}}{2014}]%
        {wang_effects_2014}
\bibfield{author}{\bibinfo{person}{Jin-Liang Wang}, \bibinfo{person}{Linda~A.
  Jackson}, \bibinfo{person}{James Gaskin}, {and} \bibinfo{person}{Hai-Zhen
  Wang}.} \bibinfo{year}{2014}\natexlab{}.
\newblock \showarticletitle{The effects of {Social} {Networking} {Site} ({SNS})
  use on college students’ friendship and well-being}.
\newblock \bibinfo{journal}{\emph{Computers in Human Behavior}}
  \bibinfo{volume}{37} (\bibinfo{date}{Aug.} \bibinfo{year}{2014}),
  \bibinfo{pages}{229--236}.
\newblock
\showISSN{0747-5632}
\urldef\tempurl%
\url{https://doi.org/10.1016/j.chb.2014.04.051}
\showDOI{\tempurl}


\bibitem[\protect\citeauthoryear{Wang, Leon, Scott, Chen, Acquisti, and
  Cranor}{Wang et~al\mbox{.}}{2013}]%
        {wang_privacy_2013}
\bibfield{author}{\bibinfo{person}{Yang Wang}, \bibinfo{person}{Pedro~Giovanni
  Leon}, \bibinfo{person}{Kevin Scott}, \bibinfo{person}{Xiaoxuan Chen},
  \bibinfo{person}{Alessandro Acquisti}, {and} \bibinfo{person}{Lorrie~Faith
  Cranor}.} \bibinfo{year}{2013}\natexlab{}.
\newblock \showarticletitle{Privacy nudges for social media: {An} exploratory
  facebook study}.
\newblock \bibinfo{journal}{\emph{WWW 2013 Companion - Proceedings of the 22nd
  International Conference on World Wide Web}}  \bibinfo{volume}{01}
  (\bibinfo{year}{2013}), \bibinfo{pages}{763--770}.
\newblock
\showISSN{9781450320382}


\bibitem[\protect\citeauthoryear{Wang, Norcie, Komanduri, Acquisti, Leon, and
  Cranor}{Wang et~al\mbox{.}}{2011}]%
        {wang_i_2011}
\bibfield{author}{\bibinfo{person}{Yang Wang}, \bibinfo{person}{Gregory
  Norcie}, \bibinfo{person}{Saranga Komanduri}, \bibinfo{person}{Alessandro
  Acquisti}, \bibinfo{person}{Pedro~Giovanni Leon}, {and}
  \bibinfo{person}{Lorrie~Faith Cranor}.} \bibinfo{year}{2011}\natexlab{}.
\newblock \showarticletitle{"{I} regretted the minute {I} pressed share": {A}
  qualitative study of regrets on {Facebook}}.
\newblock \bibinfo{journal}{\emph{SOUPS 2011 - Proceedings of the 7th Symposium
  on Usable Privacy and Security}} (\bibinfo{year}{2011}).
\newblock
\showISSN{9781450309110}
\urldef\tempurl%
\url{https://doi.org/10.1145/2078827.2078841}
\showDOI{\tempurl}


\bibitem[\protect\citeauthoryear{Zuboff}{Zuboff}{2023}]%
        {zuboff_surveillance_2023}
\bibfield{author}{\bibinfo{person}{Shoshana Zuboff}.}
  \bibinfo{year}{2023}\natexlab{}.
\newblock \showarticletitle{The age of surveillance capitalism}. In
  \bibinfo{booktitle}{\emph{Social Theory Re-Wired}}.
  \bibinfo{publisher}{Routledge}, \bibinfo{pages}{203--213}.
\newblock


\end{thebibliography}

\end{document}